\documentclass[useAMS,usenatbib,a4paper,referee]{mn2e}
\usepackage{graphicx,times}
\usepackage{amsmath}
\usepackage{longtable}

\title[GRB Redshift Distribution]
  {Cosmological Tests Using GRBs, the Star Formation Rate \\
and Possible Abundance Evolution}
  \author[Wei et al.]
    {Jun-Jie Wei$^{1,2}$, Xue-Feng Wu$^{1,3,4}$\thanks{Email:xfwu@pmo.ac.cn}, Fulvio Melia$^5$\thanks{John Woodruff Simpson Fellow},
      Da-Ming Wei$^{1,6}$, and Long-Long Feng$^{1,3}$ \\
  $^1$Purple Mountain Observatory, Chinese Academy of Sciences, Nanjing 210008, China\\
  $^2$Graduate University of Chinese Academy of Sciences, Beijing 100049, China\\
  $^3$Chinese Center for Antarctic Astronomy, Nanjing 210008, China\\
  $^4$Joint Center for Particle, Nuclear Physics and Cosmology, Nanjing University-Purple Mountain Observatory, Nanjing 210008, China\\
  $^5$Department of Physics, The Applied Math Program, and Department of Astronomy, The University of Arizona, AZ 85721, USA\\
  $^6$Key Laboratory of Dark Matter and Space Astronomy, Chinese Academy of Sciences, Nanjing, 210008, China}

\begin{document}


%

\maketitle

\begin{abstract}
The principal goal of this paper is to use attempts at reconciling the {\it Swift}
long gamma-ray bursts (LGRBs) with the star formation history (SFH) to compare the
predictions of $\Lambda$CDM with those in the $R_{\rm h}=ct$ Universe. In the
context of the former, we confirm that the latest \emph{Swift} sample of GRBs reveals an increasing
evolution in the GRB rate relative to the star formation rate (SFR) at high redshifts.
The observed discrepancy between the GRB rate and the SFR may be eliminated by assuming
a modest evolution parameterized as $(1+z)^{0.8}$---perhaps indicating a
cosmic evolution in metallicity. However, we find a higher metallicity cut of $Z=0.52Z_{\odot}$
than was seen in previous studies, which suggested that LGRBs occur preferentially in metal
poor environments, i.e., $Z\sim0.1-0.3Z_{\odot}$. We use a simple power-law approximation
to the high-\emph{z} ($\ga 3.8$) SFH, i.e., $R_{\rm SF}\propto[(1+z)/4.8]^{\alpha}$, to examine
how the high-\emph{z} SFR may be impacted by a possible abundance evolution in the
\emph{Swift} GRB sample. For an expansion history consistent with $\Lambda$CDM,
we find that the {\it Swift} redshift and luminosity distributions can be reproduced with reasonable accuracy
if $\alpha=-2.41_{-2.09}^{+1.87}$. For the $R_{\rm h}=ct$ Universe, the GRB rate is
slightly different from that in $\Lambda$CDM, but also requires an extra evolutionary
effect, with a metallicity cut of $Z=0.44Z_{\odot}$. Assuming that the SFR and GRB rate are
related via an evolving metallicity, we find that the GRB data constrain
the slope of the high-\emph{z} SFR in $R_{\rm h}=ct$ to be $\alpha=-3.60_{-2.45}^{+2.45}$.
Both cosmologies fit the GRB/SFR data rather well. However, in a one-on-one comparison
using the Aikake Information Criterion, the best-fit $R_{\rm h}=ct$ model is statistically
preferred over the best-fit $\Lambda$CDM model with a relative probability of $\sim 70\;\%$
versus $\sim 30\;\%$.
\end{abstract}

\section{Introduction}

Our understanding of the star formation history (SFH) in the Universe continues
to be refined with improving measurement techniques and a broader coverage
in redshift---now extending out to $z\ga 6$. However, direct star formation rate
(SFR) measurements are quite challenging at these high redshifts, particularly
towards the faint end of the galaxy luminosity function. Using ultraviolet and
far-infrared observations, Hopkins \& Beacom (2006) constrained the cosmic
SFH out to $z\approx6$, and found that the SFR rapidly increases at $z\la1$,
remains almost constant in the redshift range $1\la z \la4$, and then shows a
steep decline with slope $\sim -8$ at $z \ga 4$. The sharp drop at $z \ga 4$
may be due to significant dust extinction at such high redshifts. Li (2008)
derived the SFR out to $z=7.4$ by adding new ultraviolet measurements and
obtained a shallower decay ($\sim-4.46$) in this range. The high-\emph{z} SFR
has also been constrained using observations of color-selected Lyman break galaxies
(LBG; Bouwens et al. 2008; Mannucci et al. 2007; Verma et al. 2007) and Ly$\alpha$
Emitters (LAE; Ota et al. 2008). Several of the more prominent SFR determinations
are summarized in Figure~\ref{SFH} below. One can see from this plot that, due to
the inherent difficulty of making and interpreting these measurements, the various
determinations can disagree with each other even after taking the uncertainties
into account.

Gamma-ray bursts (GRBs) are the most luminous transient events in the cosmos.
Owing to their high luminosity, GRBs can be detected out to the edge of the visible
Universe, constituting a powerful tool for probing the cosmic star formation rate
from a different perspective, i.e., by studying the death rate of massive stars rather
than observing them directly during their lives. Since the successful launch of the
\emph{Swift} satellite, the number of measured GRB redshifts has increased
rapidly, and thus a reliable statistical analysis is now possible. The statistical
analysis on the GRB redshift distributions have been well investigated (e.g. Shao
et al. 2011; Robertson \& Ellis 2012; Dado \& Dar 2013). It is believed that
long bursts (LGRBs) with durations $T_{\rm 90}>2$ s (where $T_{\rm 90}$ is the time
over which $90\%$ of the prompt emission was observed; Kouveliotou et al. 1993) are
powered by the core collapse of massive stars (e.g., Woosley 1993a; Paczynski 1998;
Woosley \& Bloom 2006), an idea given strong support by several confirmed associations
between LGRBs and supernovae (Stanek et al. 2003; Hjorth et al. 2003; Chornock et al. 2010).

\begin{figure}
\includegraphics[angle=0,scale=1.0]{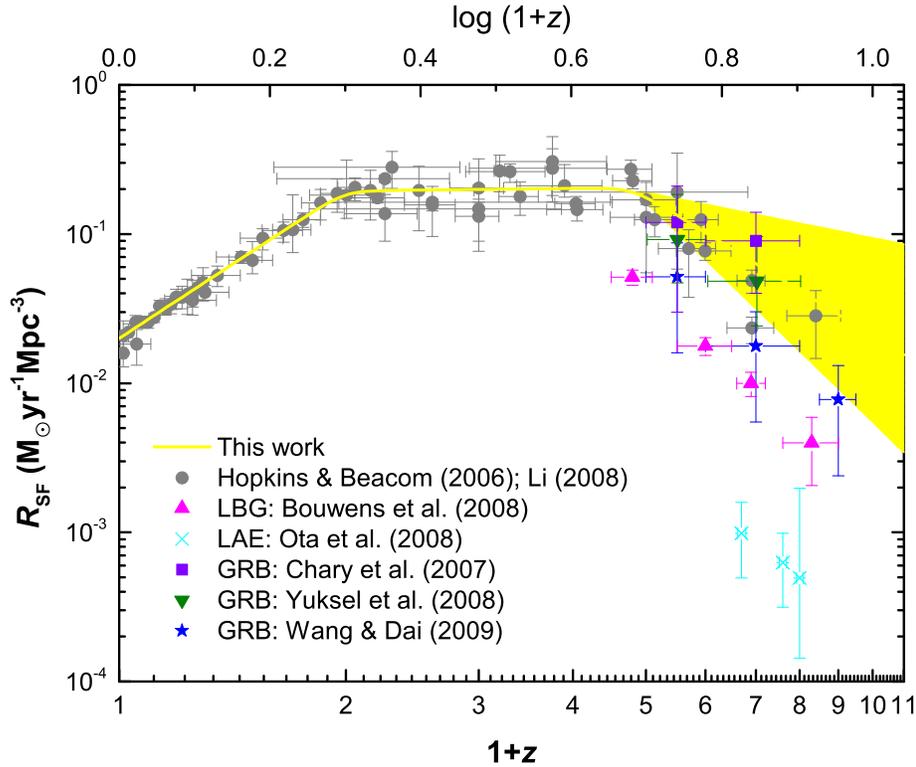}
\caption{The cosmic star formation rate as a function of redshift. The
high-\emph{z} SFR (shaded band) is constrained by the \emph{Swift}
GRB data, and is characterized by a power-law index $-5.07<\alpha<-1.05$
(see \S~4.2). Some observationally-determined SFRs are also shown
for comparison.}\label{SFH}
\end{figure}

This scenario---known as the collapsar model---suggests that the cosmic
GRB rate should in principle trace the cosmic star formation rate (Totani 1997;
Wijers et al. 1998; Blain \& Natarajan 2000; Lamb \& Reichart 2000; Porciani \&
Madau 2001; Piran 2004; Zhang \& M\'{e}sz\'{a}ros 2004; Zhang 2007).
However, observations seem to indicate that the rate of LGRBs does not
strictly follow the SFR, but instead increases with cosmic redshift faster
than the SFR, especially at high-{\em z} (Daigne et al. 2006; Le \&
Dermer 2007; Y\"{u}ksel \& Kistler et al. 2007; Salvaterra \& Chincarini 2007;
Guetta \& Piran 2007; Li 2008; Kistler et al. 2008; Salvaterra et al.
2009, 2012). This has led to the introduction of several possible mechanisms
that could produce such an observed enhancement to the GRB rate
(Daigne et al. 2006; Guetta \& Piran 2007; Le \& Dermer 2007;
Salvaterra \& Chincarini 2007; Kistler et al. 2008, 2009; Li 2008;
Salvaterra et al. 2009, 2012; Campisi et al. 2010; Qin et al. 2010;
Wanderman \& Piran 2010; Cao et al. 2011; Virgili et al. 2011; Robertson \&
Ellis 2012; Elliott et al. 2012). The idea that appears to have gained some traction
is the possibility that the difference between the GRB rate and the SFR
is due to an enhanced evolution parameterized as $(1+z)^{\delta}$
(Kistler et al. 2008), which may encompass the effects of cosmic metallicity
evolution (Langer \& Norman 2006; Li 2008), an evolution in the
stellar initial mass function (Xu \& Wei 2009; Wang \& Dai 2011), and possible
selection effects (see, e.g., Coward et al. 2008, 2013; Lu et al. 2012).

Of course, if we knew the mechanism responsible for the difference between
the GRB rate and the SFR, we could constrain the high-\emph{z} SFR very
accurately using the GRB data alone. This limitation notwithstanding, GRBs
have indeed already been used to estimate the SFR in several instances,
including the following representative cases: Chary et al. (2007) estimated
a lower limit to the SFR of $0.12\pm0.09$ and $0.09\pm0.05$
$M_{\bigodot}$ $\rm yr^{-1}$ $\rm Mpc^{-3}$ at $z=4.5$ and $6$,
respectively, using deep observations of
three $z\sim5$ GRBs with the \emph{Spitzer} Space Telescope; Y\"{u}ksel
et al. (2008) used \emph{Swift} GRB data to constrain the SFR in the range
$z=4-7$ and found that no steep drop exists in the SFR up to at least
$z\sim6.5$; Kistler et al. (2009) constrained the SFR using four years of
\emph{Swift} observations and found that the SFR to $z\ga 8$ was consistent
with LBG-based measurements; Wang \& Dai (2009) studied the high-\emph{z}
SFR up to $z\sim8.3$, but found that the SFR at $z \ga 4$ showed a steep
decay with a slope of $\sim-5.0$; and Ishida et al. (2011) used the principal
component analysis method to measure the high-\emph{z} SFR from the GRB data
and found that the level of star formation activity at $z \approx 4$ could
have been already as high as the present-day one ($\approx 0.01$
$M_{\bigodot}$ $\rm yr^{-1}$ $\rm Mpc^{-3}$).

The question of how the GRB redshift distribution is related to the SFH
is clearly still not completely answered, but there is an additional
important ingredient that has hitherto been ignored in this ongoing
discussion---the impact on this relationship from the assumed cosmological
expansion itself. Our principal goal in this paper is to update and
enlarge the GRB sample using the latest catalog of 254 {\it Swift} LGRBs
in order to carry out a comparative analysis between $\Lambda$CDM
and the $R_{\rm h}=ct$ Universe. We wish to examine the influence on
these results due to the background cosmology, and see to what
extent the implied abundance evolution depends on the expansion
scenario. We will assemble our sample in \S~2, and discuss our method of
analysis in \S~3. A possible mechanism of evolution and the implied high-\emph{z}
SFR are investigated in \S~4, together with a direct comparison between the
two cosmologies. Our discussion and conclusions are presented in \S~5.

\section{The \emph{Swift} GRB observations}

\emph{Swift} has enabled observers to greatly extend the reach of GRB measurements
relative to the pre-\emph{Swift} era, resulting in the creation of a rich data set.
To obtain reliable statistics, we consider long bursts detected by
{\em Swift} up to July, 2013, with accurate redshift measurements and durations
exceeding $T_{\rm 90}>2$ s. We calculate the isotropic-equivalent luminosity
of a GRB using
\begin{equation}
L_{\rm iso}={E_{\rm iso}(1+z)\over T_{\rm 90}}\;,
\end{equation}
where $E_{\rm iso}$ is the rest-frame isotropic equivalent $1-10^{4}$ keV gamma-ray
energy. The low-luminosity ($L_{\rm iso} < 10^{49}$ erg $\rm s^{-1}$) GRBs are not
included in our sample because they may belong to a distinct population (Soderberg
et al. 2004; Cobb et al. 2006; Liang et al. 2007; Chapman et al. 2007).

With these criteria, we combine the samples presented in Butler et al.
(2007, 2010), Perley et al. (2009), Sakamoto et al. (2011), Greiner et al. (2011),
Kr$\ddot{\rm u}$hler et al. (2011), Hjorth et al. (2012), and Perley \& Perley (2013).
For GRBs where the samples disagree, we choose the most recently measured redshifts.
The combined catalog containts 258 GRBs with known redshifts and redshift upper limits,
but four GRBs (051002, 051022, 060505, and 071112C) have incomplete fluence or burst
duration measurements and are discarded. The remaining 254 long duration GRBs with
redshifts or redshift limits serve as our base GRB catalog. Our final sample is listed
in Table 1, which includes the following information for each GRB: (1) its name;
(2) the redshift $z$; (3) the burst duration $T_{\rm 90}$; and (4) the
isotropic-equivalent energy $E_{\rm iso}$. The quantities $T_{\rm 90}$ and $E_{\rm iso}$ of 231
GRBs are directly taken from the catalog\footnote{http://butler.lab.asu.edu/Swift/index.html}
of Butler et al. (2007, 2010) and those of 14 others (050412, 050607, 050713A, 060110, 060805A,
060923A, 070521, 071011, 080319A, 080320, 080516, 081109, 081228, and 090904B) are from
Robertson \& Ellis (2012). The duration $T_{\rm 90}$ of the nine remaining
GRBs (050406, 050502B, 051016B, 060602A, 070419B, 080325, 090404, 090417B, and 090709A)
are taken from Sakamoto et al. (2011), while their isotropic energy $E_{\rm iso}$ is
calculated from the 15 to 150 keV fluences reported by Sakamoto et al. (2011); we
correct the observed fluence in a given bandpass to the cosmological rest frame
($1-10^{4}$ keV in this analysis).

\begin{center}
\begin{small}
\begin{longtable}{lcccc|lcccc}
\caption[GRB Catalog.]{GRB Catalog.} \label{GRB} \\
\hline
GRB&\emph{z}&$T_{90}$&log $E_{\rm iso}^{\Lambda{\rm CDM}}$&log ${E}_{\rm iso}^{R_{\rm h}=ct}$&GRB&\emph{z}&$T_{90}$&log $E_{\rm iso}^{\Lambda{\rm CDM}}$&log ${E}_{\rm iso}^{R_{\rm h}=ct}$\\
   &        &(s)     &(erg)&(erg)&   &   &(s)&(erg)&(erg)\\
\hline
130701A	&	1.155	&	4.62	$\pm$	0.09	&$	52.32	^{+	0.07	}_{-	0.03	}	$&$	52.23	

^{+	0.07	}_{-	0.03	}$	&	080520	&	1.545	&	2.97	$\pm$	0.24	&$	51.05	^{+	

5.78	}_{-	0.16	}	$&$	50.96	^{+	5.78	}_{-	0.16	}$	\\
130612A	&	2.006	&	6.64	$\pm$	1.06	&$	51.70	^{+	0.31	}_{-	0.09	}	$&$	51.62	

^{+	0.31	}_{-	0.09	}$	&	080516	&	$3.6^{a}$	&	$5.75^{b}$	  	  	&$	53.08	^{+	

0.22	}_{-	0.17	}	$$^{c}$&$	53.04	^{+	0.22	}_{-	0.17	}$	\\
130610A	&	2.092	&	48.45	$\pm$	2.35	&$	52.71	^{+	0.44	}_{-	0.10	}	$&$	52.63	

^{+	0.44	}_{-	0.10	}$	&	080430	&	0.767	&	16.20	$\pm$	0.78	&$	51.60	^{+	

0.34	}_{-	0.09	}	$&$	51.51	^{+	0.34	}_{-	0.09	}$	\\
130606A	&	5.913	&	278.52	$\pm$	3.54	&$	53.39	^{+	0.36	}_{-	0.08	}	$&$	53.39	

^{+	0.36	}_{-	0.08	}$	&	080413B	&	1.1	&	7.04	$\pm$	0.43	&$	52.20	^{+	

0.06	}_{-	0.05	}	$&$	52.10	^{+	0.06	}_{-	0.05	}$	\\
130604A	&	1.06	&	78.07	$\pm$	9.81	&$	51.90	^{+	0.50	}_{-	0.09	}	$&$	51.81	

^{+	0.50	}_{-	0.09	}$	&	080413A	&	2.433	&	46.62	$\pm$	0.13	&$	52.97	^{+	

0.30	}_{-	0.08	}	$&$	52.90	^{+	0.30	}_{-	0.08	}$	\\
130603B	&	0.3564	&	2.20	$\pm$	0.01	&$	50.89	^{+	0.66	}_{-	0.15	}	$&$	50.83	

^{+	0.66	}_{-	0.15	}$	&	080411	&	1.03	&	58.29	$\pm$	0.46	&$	53.38	^{+	

0.17	}_{-	0.08	}	$&$	53.28	^{+	0.17	}_{-	0.08	}$	\\
130514A	&	3.6	&	220.32	$\pm$	5.60	&$	53.60	^{+	0.12	}_{-	0.05	}	$&$	53.55	

^{+	0.12	}_{-	0.05	}$	&	080330	&	1.51	&	66.10	$\pm$	0.98	&$	51.63	^{+	

0.99	}_{-	0.06	}	$&$	51.54	^{+	0.99	}_{-	0.06	}$	\\
130511A	&	1.3033	&	4.95	$\pm$	0.82	&$	51.24	^{+	0.70	}_{-	0.14	}	$&$	51.14	

^{+	0.70	}_{-	0.14	}$	&	080325	&	$1.78^{d}$	&	$162.82^{e}$		  	&$	53.12	^{+	

0.04	}_{-	0.04	}	$$^{f
}$&$	53.03	^{+	0.04	}_{-	0.04	}$	\\
130505A	&	2.27	&	292.81	$\pm$	33.84	&$	54.31	^{+	0.45	}_{-	0.23	}	$&$	54.23	

^{+	0.45	}_{-	0.23	}$	&	080320	&	$7^{g}$	&	$13.80^{b}$	  	  	&$	53.53	^{+	

0.58	}_{-	0.07	}	$$^{c}$&$	53.56	^{+	0.58	}_{-	0.07	}$	\\
130427B	&	2.78	&	7.04	$\pm$	0.26	&$	52.50	^{+	0.39	}_{-	0.09	}	$&$	52.44	

^{+	0.39	}_{-	0.09	}$	&	080319C	&	1.95	&	32.88	$\pm$	3.27	&$	52.80	^{+	

0.37	}_{-	0.09	}	$&$	52.72	^{+	0.37	}_{-	0.09	}$	\\
130427A	&	0.3399	&	324.70	$\pm$	2.50	&$	53.66	^{+	0.19	}_{-	0.11	}	$&$	53.60	

^{+	0.19	}_{-	0.11	}$	&	080319B	&	0.937	&	147.32	$\pm$	2.50	&$	54.58	^{+	

0.26	}_{-	0.17	}	$&$	54.49	^{+	0.26	}_{-	0.17	}$	\\
130420A	&	1.297	&	114.84	$\pm$	4.84	&$	52.72	^{+	0.07	}_{-	0.05	}	$&$	52.63	

^{+	0.07	}_{-	0.05	}$	&	080319A	&	$2.2^{g}$	&	$43.60^{b}$	  	  	&$	53.47	^{+	

0.38	}_{-	0.06	}	$$^{c}$&$	53.39	^{+	0.38	}_{-	0.06	}$	\\
130418A	&	1.217	&	97.92	$\pm$	2.26	&$	51.77	^{+	0.14	}_{-	0.08	}	$&$	51.68	

^{+	0.14	}_{-	0.08	}$	&	080310	&	2.4266	&	361.92	$\pm$	3.75	&$	52.78	^{+	

0.78	}_{-	0.07	}	$&$	52.71	^{+	0.78	}_{-	0.07	}$	\\
130408A	&	3.758	&	5.64	$\pm$	0.31	&$	53.08	^{+	0.55	}_{-	0.13	}	$&$	53.04	

^{+	0.55	}_{-	0.13	}$	&	080210	&	2.641	&	43.89	$\pm$	4.36	&$	52.72	^{+	

0.39	}_{-	0.08	}	$&$	52.65	^{+	0.39	}_{-	0.08	}$	\\
130215A	&	0.597	&	89.05	$\pm$	8.39	&$	51.89	^{+	0.31	}_{-	0.07	}	$&$	51.81	

^{+	0.31	}_{-	0.07	}$	&	080207	&	2.0858	&	310.98	$\pm$	9.34	&$	53.05	^{+	

0.23	}_{-	0.07	}	$&$	52.96	^{+	0.23	}_{-	0.07	}$	\\
130131B	&	2.539	&	4.74	$\pm$	0.21	&$	52.23	^{+	0.03	}_{-	0.03	}	$&$	52.16	

^{+	0.03	}_{-	0.03	}$	&	080129	&	4.349	&	45.60	$\pm$	3.00	&$	52.90	^{+	

0.42	}_{-	0.20	}	$&$	52.87	^{+	0.42	}_{-	0.19	}$	\\
121229A	&	2.707	&	26.64	$\pm$	2.15	&$	51.85	^{+	0.95	}_{-	0.10	}	$&$	51.79	

^{+	0.95	}_{-	0.10	}$	&	071227	&	0.383	&	2.20	$\pm$	0.16	&$	50.45	^{+	

0.60	}_{-	0.22	}	$&$	50.39	^{+	0.60	}_{-	0.22	}$	\\
121211A	&	1.023	&	184.14	$\pm$	2.31	&$	51.80	^{+	0.65	}_{-	0.09	}	$&$	51.70	

^{+	0.65	}_{-	0.09	}$	&	071122	&	1.14	&	79.20	$\pm$	4.88	&$	51.55	^{+	

0.64	}_{-	0.14	}	$&$	51.46	^{+	0.64	}_{-	0.14	}$	\\
121201A	&	3.385	&	39.04	$\pm$	2.93	&$	52.39	^{+	0.38	}_{-	0.08	}	$&$	52.34	

^{+	0.38	}_{-	0.08	}$	&	071117	&	1.331	&	6.48	$\pm$	0.76	&$	52.29	^{+	

0.18	}_{-	0.07	}	$&$	52.20	^{+	0.18	}_{-	0.07	}$	\\
121128A	&	2.2	&	25.65	$\pm$	5.47	&$	52.98	^{+	0.10	}_{-	0.07	}	$&$	52.91	

^{+	0.10	}_{-	0.07	}$	&	071031	&	2.692	&	187.18	$\pm$	7.12	&$	52.61	^{+	

0.45	}_{-	0.07	}	$&$	52.54	^{+	0.45	}_{-	0.07	}$	\\
121027A	&	1.77	&	69.30	$\pm$	1.90	&$	52.39	^{+	0.11	}_{-	0.09	}	$&$	52.31	

^{+	0.11	}_{-	0.09	}$	&	071021	&	2.145	&	204.96	$\pm$	17.95	&$	53.00	^{+	

0.43	}_{-	0.14	}	$&$	52.92	^{+	0.43	}_{-	0.14	}$	\\
121024A	&	2.298	&	12.46	$\pm$	0.39	&$	52.40	^{+	0.38	}_{-	0.16	}	$&$	52.32	

^{+	0.38	}_{-	0.16	}$	&	071020	&	2.145	&	4.40	$\pm$	0.27	&$	53.00	^{+	

0.43	}_{-	0.14	}	$&$	52.92	^{+	0.43	}_{-	0.14	}$	\\
120922A	&	3.1	&	179.54	$\pm$	6.27	&$	53.28	^{+	0.21	}_{-	0.04	}	$&$	53.22	

^{+	0.21	}_{-	0.04	}$	&	071011	&	$5^{g}$	&	$80.90^{b}$	  	  	&$	54.37	^{+	

0.34	}_{-	0.19	}	$$^{c}$&$	54.36	^{+	0.34	}_{-	0.19	}$	\\
120909A	&	3.93	&	617.70	$\pm$	30.95	&$	53.68	^{+	0.48	}_{-	0.09	}	$&$	53.64	

^{+	0.48	}_{-	0.09	}$	&	071010B	&	0.947	&	34.68	$\pm$	1.02	&$	52.26	^{+	

0.09	}_{-	0.03	}	$&$	52.16	^{+	0.09	}_{-	0.03	}$	\\
120907A	&	0.97	&	6.27	$\pm$	0.28	&$	51.29	^{+	0.40	}_{-	0.05	}	$&$	51.20	

^{+	0.40	}_{-	0.05	}$	&	071010A	&	0.98	&	22.40	$\pm$	1.70	&$	51.13	^{+	

0.81	}_{-	0.07	}	$&$	51.04	^{+	0.81	}_{-	0.07	}$	\\
120815A	&	2.358	&	9.68	$\pm$	1.21	&$	52.01	^{+	0.90	}_{-	0.09	}	$&$	51.94	

^{+	0.90	}_{-	0.09	}$	&	071003	&	1.605	&	148.32	$\pm$	0.68	&$	53.27	^{+	

0.35	}_{-	0.15	}	$&$	53.17	^{+	0.35	}_{-	0.15	}$	\\
120811C	&	2.671	&	25.20	$\pm$	1.26	&$	52.88	^{+	0.02	}_{-	0.10	}	$&$	52.81	

^{+	0.02	}_{-	0.10	}$	&	070810A	&	2.17	&	7.68	$\pm$	0.41	&$	51.97	^{+	

0.13	}_{-	0.05	}	$&$	51.89	^{+	0.13	}_{-	0.05	}$	\\
120802A	&	3.796	&	50.16	$\pm$	1.52	&$	52.83	^{+	0.09	}_{-	0.07	}	$&$	52.79	

^{+	0.09	}_{-	0.07	}$	&	070802	&	2.45	&	14.72	$\pm$	0.61	&$	51.71	^{+	

0.46	}_{-	0.08	}	$&$	51.63	^{+	0.46	}_{-	0.08	}$	\\
120729A	&	0.8	&	78.65	$\pm$	6.50	&$	51.86	^{+	0.40	}_{-	0.08	}	$&$	51.77	

^{+	0.40	}_{-	0.08	}$	&	070721B	&	3.626	&	330.66	$\pm$	6.28	&$	53.51	^{+	

0.32	}_{-	0.19	}	$&$	53.47	^{+	0.32	}_{-	0.19	}$	\\
120724A	&	1.48	&	49.17	$\pm$	4.33	&$	51.78	^{+	0.65	}_{-	0.10	}	$&$	51.68	

^{+	0.65	}_{-	0.10	}$	&	070714B	&	0.92	&	64.18	$\pm$	1.60	&$	51.50	^{+	

0.60	}_{-	0.15	}	$&$	51.41	^{+	0.60	}_{-	0.15	}$	\\
120722A	&	0.9586	&	37.31	$\pm$	2.46	&$	51.68	^{+	0.71	}_{-	0.03	}	$&$	51.59	

^{+	0.71	}_{-	0.03	}$	&	070612A	&	0.617	&	254.74	$\pm$	3.63	&$	52.30	^{+	

0.40	}_{-	0.09	}	$&$	52.22	^{+	0.40	}_{-	0.09	}$	\\
120712A	&	4.1745	&	18.46	$\pm$	1.08	&$	52.97	^{+	0.19	}_{-	0.07	}	$&$	52.94	

^{+	0.19	}_{-	0.07	}$	&	070611	&	2.04	&	11.31	$\pm$	0.45	&$	51.72	^{+	

0.30	}_{-	0.10	}	$&$	51.64	^{+	0.30	}_{-	0.10	}$	\\
120404A	&	2.876	&	40.50	$\pm$	1.49	&$	52.65	^{+	0.30	}_{-	0.08	}	$&$	52.58	

^{+	0.30	}_{-	0.08	}$	&	070529	&	2.4996	&	112.21	$\pm$	2.94	&$	52.98	^{+	

0.40	}_{-	0.16	}	$&$	52.91	^{+	0.40	}_{-	0.16	}$	\\
120327A	&	2.813	&	71.20	$\pm$	2.33	&$	53.00	^{+	0.17	}_{-	0.05	}	$&$	52.93	

^{+	0.17	}_{-	0.05	}$	&	070521	&	$1.35^{h}$	&	$38.60^{b}$	  	  	&$	53.40	^{+	

0.38	}_{-	0.15	}	$$^{c}$&$	53.31	^{+	0.38	}_{-	0.15	}$	\\
120326A	&	1.798	&	72.72	$\pm$	3.08	&$	52.49	^{+	0.07	}_{-	0.03	}	$&$	52.40	

^{+	0.07	}_{-	0.03	}$	&	070518	&	1.16	&	5.34	$\pm$	0.19	&$	50.94	^{+	

0.75	}_{-	0.06	}	$&$	50.85	^{+	0.75	}_{-	0.06	}$	\\
120119A	&	1.728	&	70.40	$\pm$	4.32	&$	53.33	^{+	0.08	}_{-	0.04	}	$&$	53.24	

^{+	0.08	}_{-	0.04	}$	&	070508	&	0.82	&	21.20	$\pm$	0.25	&$	52.90	^{+	

0.09	}_{-	0.06	}	$&$	52.81	^{+	0.09	}_{-	0.06	}$	\\
120118B	&	2.943	&	30.78	$\pm$	2.85	&$	52.81	^{+	0.55	}_{-	0.04	}	$&$	52.75	

^{+	0.55	}_{-	0.04	}$	&	070506	&	2.31	&	3.55	$\pm$	0.17	&$	51.42	^{+	

0.28	}_{-	0.09	}	$&$	51.35	^{+	0.28	}_{-	0.09	}$	\\
111229A	&	1.3805	&	2.79	$\pm$	0.25	&$	50.97	^{+	0.72	}_{-	0.06	}	$&$	50.88	

^{+	0.72	}_{-	0.06	}$	&	070419B	&	$1.9591^{i}$	&	$238.14^{e}$		  	&$	53.38	^{+	

0.01	}_{-	0.01	}	$$^{f}$&$	53.29	^{+	0.01	}_{-	0.01	}$	\\
111228A	&	0.714	&	101.40	$\pm$	1.31	&$	52.56	^{+	0.11	}_{-	0.10	}	$&$	52.48	

^{+	0.11	}_{-	0.10	}$	&	070419A	&	0.97	&	161.25	$\pm$	8.87	&$	51.39	^{+	

0.42	}_{-	0.09	}	$&$	51.29	^{+	0.42	}_{-	0.09	}$	\\
111123A	&	3.1516	&	235.20	$\pm$	6.58	&$	53.39	^{+	0.15	}_{-	0.07	}	$&$	53.34	

^{+	0.15	}_{-	0.07	}$	&	070411	&	2.954	&	108.56	$\pm$	3.62	&$	53.02	^{+	

0.34	}_{-	0.08	}	$&$	52.96	^{+	0.34	}_{-	0.08	}$	\\
111209A	&	0.677	&	4.64	$\pm$	0.33	&$	51.18	^{+	0.77	}_{-	0.17	}	$&$	51.09	

^{+	0.77	}_{-	0.17	}$	&	070318	&	0.836	&	51.00	$\pm$	2.32	&$	51.98	^{+	

0.41	}_{-	0.10	}	$&$	51.89	^{+	0.41	}_{-	0.10	}$	\\
111107A	&	2.893	&	31.59	$\pm$	2.44	&$	52.52	^{+	0.44	}_{-	0.11	}	$&$	52.46	

^{+	0.44	}_{-	0.11	}$	&	070306	&	1.497	&	261.36	$\pm$	6.65	&$	52.80	^{+	

0.39	}_{-	0.08	}	$&$	52.71	^{+	0.39	}_{-	0.08	}$	\\
111008A	&	4.9898	&	75.66	$\pm$	2.25	&$	53.69	^{+	0.34	}_{-	0.06	}	$&$	53.68	

^{+	0.34	}_{-	0.06	}$	&	070208	&	1.165	&	52.48	$\pm$	0.85	&$	51.47	^{+	

0.34	}_{-	0.13	}	$&$	51.37	^{+	0.34	}_{-	0.13	}$	\\
110818A	&	3.36	&	77.28	$\pm$	5.61	&$	53.16	^{+	0.40	}_{-	0.07	}	$&$	53.11	

^{+	0.40	}_{-	0.07	}$	&	070129	&	2.3384	&	92.15	$\pm$	2.24	&$	52.49	^{+	

0.11	}_{-	0.09	}	$&$	52.41	^{+	0.11	}_{-	0.09	}$	\\
110808A	&	1.348	&	39.38	$\pm$	3.44	&$	51.45	^{+	0.91	}_{-	0.09	}	$&$	51.36	

^{+	0.91	}_{-	0.09	}$	&	070110	&	2.352	&	47.70	$\pm$	1.54	&$	52.45	^{+	

0.30	}_{-	0.08	}	$&$	52.38	^{+	0.30	}_{-	0.08	}$	\\
110801A	&	1.858	&	400.40	$\pm$	1.99	&$	52.80	^{+	0.19	}_{-	0.09	}	$&$	52.72	

^{+	0.19	}_{-	0.09	}$	&	070103	&	2.6208	&	10.92	$\pm$	0.14	&$	51.70	^{+	

0.47	}_{-	0.09	}	$&$	51.63	^{+	0.47	}_{-	0.09	}$	\\
110731A	&	2.83	&	46.56	$\pm$	7.14	&$	53.56	^{+	0.32	}_{-	0.14	}	$&$	53.50	

^{+	0.32	}_{-	0.14	}$	&	061222B	&	3.355	&	42.00	$\pm$	2.15	&$	52.92	^{+	

0.39	}_{-	0.08	}	$&$	52.87	^{+	0.39	}_{-	0.08	}$	\\
110715A	&	0.82	&	13.15	$\pm$	1.40	&$	52.48	^{+	0.04	}_{-	0.03	}	$&$	52.39	

^{+	0.04	}_{-	0.03	}$	&	061222A	&	2.088	&	81.65	$\pm$	4.24	&$	53.32	^{+	

0.25	}_{-	0.07	}	$&$	53.24	^{+	0.25	}_{-	0.07	}$	\\
110503A	&	1.613	&	9.31	$\pm$	0.64	&$	53.07	^{+	0.16	}_{-	0.08	}	$&$	52.98	

^{+	0.16	}_{-	0.08	}$	&	061126	&	1.159	&	26.78	$\pm$	0.46	&$	52.89	^{+	

0.39	}_{-	0.14	}	$&$	52.80	^{+	0.39	}_{-	0.14	}$	\\
110422A	&	1.77	&	26.73	$\pm$	0.29	&$	53.65	^{+	0.03	}_{-	0.02	}	$&$	53.57	

^{+	0.03	}_{-	0.02	}$	&	061121	&	1.314	&	83.00	$\pm$	12.50	&$	53.30	^{+	

0.24	}_{-	0.11	}	$&$	53.20	^{+	0.24	}_{-	0.11	}$	\\
110213A	&	1.46	&	43.12	$\pm$	3.47	&$	52.72	^{+	0.26	}_{-	0.08	}	$&$	52.62	

^{+	0.26	}_{-	0.08	}$	&	061110B	&	3.44	&	32.39	$\pm$	0.45	&$	53.12	^{+	

0.37	}_{-	0.26	}	$&$	53.07	^{+	0.37	}_{-	0.26	}$	\\
110205A	&	2.22	&	277.02	$\pm$	4.67	&$	53.48	^{+	0.10	}_{-	0.04	}	$&$	53.41	

^{+	0.10	}_{-	0.04	}$	&	061110A	&	0.757	&	47.04	$\pm$	1.80	&$	51.46	^{+	

0.43	}_{-	0.09	}	$&$	51.38	^{+	0.43	}_{-	0.09	}$	\\
110128A	&	2.339	&	17.10	$\pm$	0.70	&$	52.36	^{+	0.49	}_{-	0.22	}	$&$	52.28	

^{+	0.49	}_{-	0.22	}$	&	061021	&	0.3463	&	12.06	$\pm$	0.32	&$	51.40	^{+	

0.38	}_{-	0.15	}	$&$	51.34	^{+	0.38	}_{-	0.15	}$	\\
101225A	&	0.847	&	63.00	$\pm$	6.97	&$	51.43	^{+	0.64	}_{-	0.33	}	$&$	51.34	

^{+	0.64	}_{-	0.33	}$	&	061007	&	1.261	&	74.90	$\pm$	0.51	&$	54.17	^{+	

0.33	}_{-	0.17	}	$&$	54.08	^{+	0.33	}_{-	0.17	}$	\\
101219B	&	0.55	&	41.80	$\pm$	1.45	&$	51.47	^{+	0.52	}_{-	0.08	}	$&$	51.39	

^{+	0.52	}_{-	0.08	}$	&	060927	&	5.4636	&	23.03	$\pm$	0.26	&$	52.95	^{+	

0.10	}_{-	0.06	}	$&$	52.95	^{+	0.10	}_{-	0.06	}$	\\
101213A	&	0.414	&	175.68	$\pm$	15.30	&$	51.85	^{+	0.32	}_{-	0.17	}	$&$	51.78	

^{+	0.32	}_{-	0.17	}$	&	060926	&	3.2	&	7.05	$\pm$	0.39	&$	51.95	^{+	

1.13	}_{-	0.08	}	$&$	51.90	^{+	1.13	}_{-	0.08	}$	\\
100906A	&	1.727	&	116.85	$\pm$	0.69	&$	53.14	^{+	0.21	}_{-	0.07	}	$&$	53.05	

^{+	0.21	}_{-	0.07	}$	&	060923A	&	$4^{g}$	&	$51.50^{b}$	  	  	&$	53.30	^{+	

0.20	}_{-	0.10	}	$$^{c}$&$	53.27	^{+	0.20	}_{-	0.10	}$	\\
100901A	&	1.408	&	459.19	$\pm$	10.66	&$	52.26	^{+	0.57	}_{-	0.12	}	$&$	52.17	

^{+	0.57	}_{-	0.12	}$	&	060912A	&	0.937	&	5.92	$\pm$	0.35	&$	51.92	^{+	

0.26	}_{-	0.12	}	$&$	51.83	^{+	0.26	}_{-	0.12	}$	\\
100816A	&	0.8049	&	2.50	$\pm$	0.22	&$	51.75	^{+	0.15	}_{-	0.06	}	$&$	51.66	

^{+	0.15	}_{-	0.06	}$	&	060908	&	1.8836	&	18.48	$\pm$	0.17	&$	52.61	^{+	

0.18	}_{-	0.07	}	$&$	52.53	^{+	0.18	}_{-	0.07	}$	\\
100814A	&	1.44	&	176.96	$\pm$	3.61	&$	52.79	^{+	0.16	}_{-	0.05	}	$&$	52.70	

^{+	0.16	}_{-	0.05	}$	&	060906	&	3.686	&	72.96	$\pm$	9.41	&$	53.11	^{+	

0.43	}_{-	0.04	}	$&$	53.07	^{+	0.43	}_{-	0.04	}$	\\
100728B	&	2.106	&	11.52	$\pm$	0.78	&$	52.39	^{+	0.33	}_{-	0.07	}	$&$	52.31	

^{+	0.33	}_{-	0.07	}$	&	060904B	&	0.703	&	171.04	$\pm$	2.29	&$	51.49	^{+	

0.28	}_{-	0.09	}	$&$	51.40	^{+	0.28	}_{-	0.09	}$	\\
100728A	&	1.567	&	222.00	$\pm$	6.89	&$	53.82	^{+	0.14	}_{-	0.08	}	$&$	53.73	

^{+	0.14	}_{-	0.08	}$	&	060814	&	0.84	&	159.16	$\pm$	4.08	&$	52.95	^{+	

0.03	}_{-	0.18	}	$&$	52.86	^{+	0.03	}_{-	0.18	}$	\\
100621A	&	0.542	&	66.33	$\pm$	1.27	&$	52.46	^{+	0.05	}_{-	0.03	}	$&$	52.38	

^{+	0.05	}_{-	0.03	}$	&	060805A	&	$3.8^{g}$	&	$4.93^{b}$	  	  	&$	52.26	^{+	

0.65	}_{-	0.12	}	$$^{c}$&$	52.22	^{+	0.65	}_{-	0.12	}$	\\
100615A	&	1.398	&	43.46	$\pm$	1.30	&$	52.62	^{+	0.08	}_{-	0.05	}	$&$	52.53	

^{+	0.08	}_{-	0.05	}$	&	060729	&	0.54	&	119.14	$\pm$	1.40	&$	51.49	^{+	

0.33	}_{-	0.08	}	$&$	51.41	^{+	0.33	}_{-	0.08	}$	\\
100513A	&	4.772	&	65.10	$\pm$	4.39	&$	52.92	^{+	0.37	}_{-	0.08	}	$&$	52.90	

^{+	0.37	}_{-	0.08	}$	&	060719	&	1.532	&	57.00	$\pm$	0.84	&$	52.16	^{+	

0.55	}_{-	0.03	}	$&$	52.07	^{+	0.55	}_{-	0.03	}$	\\
100425A	&	1.755	&	43.56	$\pm$	1.03	&$	51.81	^{+	0.73	}_{-	0.12	}	$&$	51.72	

^{+	0.73	}_{-	0.12	}$	&	060714	&	2.711	&	118.72	$\pm$	1.87	&$	52.90	^{+	

0.42	}_{-	0.05	}	$&$	52.83	^{+	0.42	}_{-	0.05	}$	\\
100424A	&	2.465	&	110.25	$\pm$	5.30	&$	52.50	^{+	0.30	}_{-	0.08	}	$&$	52.42	

^{+	0.30	}_{-	0.08	}$	&	060708	&	1.92	&	7.50	$\pm$	0.45	&$	51.78	^{+	

0.20	}_{-	0.07	}	$&$	51.70	^{+	0.20	}_{-	0.07	}$	\\
100418A	&	0.624	&	9.63	$\pm$	0.81	&$	50.73	^{+	0.77	}_{-	0.04	}	$&$	50.65	

^{+	0.77	}_{-	0.04	}$	&	060707	&	3.425	&	75.14	$\pm$	2.46	&$	52.80	^{+	

0.14	}_{-	0.07	}	$&$	52.75	^{+	0.14	}_{-	0.07	}$	\\
100316B	&	1.18	&	4.30	$\pm$	0.34	&$	51.08	^{+	0.86	}_{-	0.03	}	$&$	50.99	

^{+	0.86	}_{-	0.03	}$	&	060614	&	0.125	&	108.80	$\pm$	0.86	&$	51.40	^{+	

0.07	}_{-	0.08	}	$&$	51.37	^{+	0.07	}_{-	0.08	}$	\\
100302A	&	4.813	&	31.72	$\pm$	3.11	&$	52.36	^{+	0.72	}_{-	0.04	}	$&$	52.35	

^{+	0.72	}_{-	0.04	}$	&	060607A	&	3.082	&	102.55	$\pm$	3.35	&$	52.97	^{+	

0.32	}_{-	0.08	}	$&$	52.91	^{+	0.32	}_{-	0.08	}$	\\
100219A	&	4.667	&	31.05	$\pm$	2.84	&$	52.46	^{+	0.55	}_{-	0.13	}	$&$	52.44	

^{+	0.55	}_{-	0.13	}$	&	060605	&	3.78	&	18.54	$\pm$	1.16	&$	52.34	^{+	

0.53	}_{-	0.10	}	$&$	52.30	^{+	0.53	}_{-	0.10	}$	\\
091208B	&	1.063	&	15.21	$\pm$	1.31	&$	52.16	^{+	0.17	}_{-	0.07	}	$&$	52.06	

^{+	0.17	}_{-	0.07	}$	&	060604	&	2.1357	&	39.90	$\pm$	0.70	&$	51.73	^{+	

0.96	}_{-	0.10	}	$&$	51.65	^{+	0.96	}_{-	0.10	}$	\\
091127	&	0.49	&	9.57	$\pm$	0.56	&$	52.16	^{+	0.31	}_{-	0.02	}	$&$	52.09	

^{+	0.31	}_{-	0.02	}$	&	060602A	&	$0.787^{i}$	&	$74.68^{e}$		  	&$	51.98	^{+	

0.04	}_{-	0.04	}	$$^{f}$&$	51.89	^{+	0.04	}_{-	0.04	}$	\\
091109A	&	3.076	&	49.68	$\pm$	4.60	&$	53.13	^{+	0.31	}_{-	0.22	}	$&$	53.08	

^{+	0.31	}_{-	0.22	}$	&	060526	&	3.221	&	295.55	$\pm$	4.01	&$	52.73	^{+	

0.47	}_{-	0.03	}	$&$	52.68	^{+	0.47	}_{-	0.03	}$	\\
091029	&	2.752	&	39.96	$\pm$	1.28	&$	52.91	^{+	0.06	}_{-	0.07	}	$&$	52.85	

^{+	0.06	}_{-	0.07	}$	&	060522	&	5.11	&	74.10	$\pm$	2.30	&$	52.87	^{+	

0.40	}_{-	0.08	}	$&$	52.86	^{+	0.40	}_{-	0.08	}$	\\
091024	&	1.092	&	114.73	$\pm$	4.95	&$	52.80	^{+	0.37	}_{-	0.15	}	$&$	52.70	

^{+	0.37	}_{-	0.15	}$	&	060512	&	0.4428	&	8.37	$\pm$	0.36	&$	50.31	^{+	

0.65	}_{-	0.09	}	$&$	50.24	^{+	0.65	}_{-	0.09	}$	\\
091020	&	1.71	&	39.00	$\pm$	1.07	&$	52.67	^{+	0.30	}_{-	0.08	}	$&$	52.58	

^{+	0.30	}_{-	0.08	}$	&	060510B	&	4.9	&	229.89	$\pm$	2.77	&$	53.37	^{+	

0.19	}_{-	0.08	}	$&$	53.36	^{+	0.19	}_{-	0.08	}$	\\
091018	&	0.971	&	4.44	$\pm$	0.15	&$	51.82	^{+	0.10	}_{-	0.05	}	$&$	51.72	

^{+	0.10	}_{-	0.05	}$	&	060502A	&	1.51	&	30.24	$\pm$	4.18	&$	52.47	^{+	

0.39	}_{-	0.10	}	$&$	52.38	^{+	0.39	}_{-	0.10	}$	\\
090927	&	1.37	&	18.36	$\pm$	1.33	&$	51.35	^{+	0.71	}_{-	0.07	}	$&$	51.26	

^{+	0.71	}_{-	0.07	}$	&	060428B	&	0.348	&	20.46	$\pm$	0.62	&$	50.31	^{+	

0.28	}_{-	0.10	}	$&$	50.25	^{+	0.28	}_{-	0.10	}$	\\
090926B	&	1.24	&	126.36	$\pm$	5.21	&$	52.56	^{+	0.06	}_{-	0.03	}	$&$	52.47	

^{+	0.06	}_{-	0.03	}$	&	060418	&	1.489	&	103.24	$\pm$	10.33	&$	52.93	^{+	

0.28	}_{-	0.06	}	$&$	52.84	^{+	0.28	}_{-	0.06	}$	\\
090904B	&	$5^{j}$	&	$64.00^{b}$	  	  	&$	53.54	^{+	0.18	}_{-	0.18	}	$$^{c}$&$	53.53	

^{+	0.18	}_{-	0.18	}$	&	060306	&	3.5	&	60.96	$\pm$	0.80	&$	52.88	^{+	

0.15	}_{-	0.06	}	$&$	52.84	^{+	0.15	}_{-	0.06	}$	\\
090814A	&	0.696	&	113.16	$\pm$	12.99	&$	51.39	^{+	0.24	}_{-	0.08	}	$&$	51.30	

^{+	0.24	}_{-	0.08	}$	&	060223A	&	4.41	&	8.40	$\pm$	0.28	&$	52.50	^{+	

0.17	}_{-	0.07	}	$&$	52.48	^{+	0.17	}_{-	0.07	}$	\\
090812	&	2.452	&	99.76	$\pm$	15.30	&$	53.32	^{+	0.38	}_{-	0.12	}	$&$	53.25	

^{+	0.38	}_{-	0.12	}$	&	060210	&	3.91	&	369.94	$\pm$	20.65	&$	53.63	^{+	

0.36	}_{-	0.08	}	$&$	53.59	^{+	0.36	}_{-	0.08	}$	\\
090809	&	2.737	&	192.92	$\pm$	5.24	&$	52.16	^{+	0.74	}_{-	0.13	}	$&$	52.09	

^{+	0.74	}_{-	0.13	}$	&	060206	&	4.045	&	6.06	$\pm$	0.16	&$	52.63	^{+	

0.12	}_{-	0.07	}	$&$	52.60	^{+	0.12	}_{-	0.07	}$	\\
090726	&	2.71	&	51.03	$\pm$	0.97	&$	52.27	^{+	0.49	}_{-	0.10	}	$&$	52.21	

^{+	0.49	}_{-	0.10	}$	&	060202	&	0.783	&	205.92	$\pm$	2.52	&$	51.83	^{+	

0.41	}_{-	0.07	}	$&$	51.74	^{+	0.41	}_{-	0.07	}$	\\
090715B	&	3	&	267.54	$\pm$	4.54	&$	53.39	^{+	0.28	}_{-	0.09	}	$&$	53.33	

^{+	0.28	}_{-	0.09	}$	&	060124	&	2.296	&	8.16	$\pm$	0.19	&$	51.84	^{+	

0.44	}_{-	0.10	}	$&$	51.76	^{+	0.44	}_{-	0.10	}$	\\
090709A	&	$1.8^{d}$	&	$88.73^{e}$	  	  	&$	52.61	^{+	0.05	}_{-	0.05	}	$$^{f}$&$	52.52	

^{+	0.05	}_{-	0.05	}$	&	060116	&	6.6	&	36.00	$\pm$	1.21	&$	53.30	^{+	

0.38	}_{-	0.12	}	$&$	53.32	^{+	0.38	}_{-	0.12	}$	\\
090618	&	0.54	&	115.20	$\pm$	0.43	&$	53.17	^{+	0.04	}_{-	0.03	}	$&$	53.10	

^{+	0.04	}_{-	0.03	}$	&	060115	&	3.53	&	109.89	$\pm$	1.14	&$	52.79	^{+	

0.17	}_{-	0.07	}	$&$	52.75	^{+	0.17	}_{-	0.07	}$	\\
090529	&	2.625	&	79.79	$\pm$	3.52	&$	52.41	^{+	0.24	}_{-	0.09	}	$&$	52.34	

^{+	0.24	}_{-	0.09	}$	&	060110	&	$5^{g}$	&	$21.10^{b}$	  	  	&$	53.92	^{+	

0.35	}_{-	0.08	}	$$^{c}$&$	53.91	^{+	0.35	}_{-	0.08	}$	\\
090519	&	3.85	&	81.77	$\pm$	6.00	&$	53.18	^{+	0.38	}_{-	0.24	}	$&$	53.14	

^{+	0.38	}_{-	0.24	}$	&	060108	&	2.03	&	15.28	$\pm$	1.10	&$	51.78	^{+	

0.62	}_{-	0.06	}	$&$	51.70	^{+	0.62	}_{-	0.06	}$	\\
090516	&	4.109	&	228.48	$\pm$	9.45	&$	53.73	^{+	0.38	}_{-	0.10	}	$&$	53.69	

^{+	0.38	}_{-	0.10	}$	&	051227	&	0.714	&	4.30	$\pm$	0.19	&$	50.90	^{+	

0.57	}_{-	0.23	}	$&$	50.81	^{+	0.57	}_{-	0.23	}$	\\
090429B	&	9.4	&	5.80	$\pm$	0.29	&$	52.74	^{+	0.13	}_{-	0.07	}	$&$	52.81	

^{+	0.13	}_{-	0.07	}$	&	051117B	&	0.481	&	10.45	$\pm$	0.25	&$	50.23	^{+	

0.56	}_{-	0.11	}	$&$	50.16	^{+	0.56	}_{-	0.11	}$	\\
090424	&	0.544	&	50.28	$\pm$	0.53	&$	52.43	^{+	0.06	}_{-	0.05	}	$&$	52.36	

^{+	0.06	}_{-	0.05	}$	&	051111	&	1.55	&	50.96	$\pm$	2.45	&$	52.70	^{+	

0.33	}_{-	0.09	}	$&$	52.61	^{+	0.33	}_{-	0.09	}$	\\
090423	&	8.26	&	12.36	$\pm$	0.59	&$	52.93	^{+	0.09	}_{-	0.07	}	$&$	52.98	

^{+	0.09	}_{-	0.07	}$	&	051109A	&	2.346	&	4.90	$\pm$	0.30	&$	52.35	^{+	

0.49	}_{-	0.08	}	$&$	52.28	^{+	0.49	}_{-	0.08	}$	\\
090418	&	1.608	&	57.97	$\pm$	0.85	&$	52.95	^{+	0.31	}_{-	0.15	}	$&$	52.86	

^{+	0.31	}_{-	0.15	}$	&	051016B	&	$0.9364^{i}$	&	$4.02^{e}$		  	&$	51.15	^{+	

0.06	}_{-	0.06	}	$$^{f}$&$	51.06	^{+	0.06	}_{-	0.06	}$	\\
090417B	&	$0.345^{d}$	&	$282.49^{e}$	  	  	&$	51.41	^{+	0.03	}_{-	0.03	}	$$^{f}$&$	51.35	

^{+	0.03	}_{-	0.03	}$	&	051006	&	1.059	&	26.46	$\pm$	0.53	&$	52.02	^{+	

0.34	}_{-	0.20	}	$&$	51.93	^{+	0.34	}_{-	0.20	}$	\\
090407	&	1.4485	&	147.52	$\pm$	1.02	&$	51.71	^{+	0.74	}_{-	0.14	}	$&$	51.62	

^{+	0.74	}_{-	0.14	}$	&	051001	&	2.4296	&	55.90	$\pm$	1.63	&$	52.38	^{+	

0.07	}_{-	0.11	}	$&$	52.31	^{+	0.07	}_{-	0.11	}$	\\
090404	&	$3^{d}$	&	$82.01^{e}$	  	  	&$	53.30	^{+	0.02	}_{-	0.02	}	$$^{f}$&$	53.24	

^{+	0.02	}_{-	0.02	}$	&	050922C	&	2.198	&	4.56	$\pm$	0.12	&$	52.60	^{+	

0.30	}_{-	0.08	}	$&$	52.52	^{+	0.30	}_{-	0.08	}$	\\
090313	&	3.375	&	90.24	$\pm$	6.75	&$	52.67	^{+	0.67	}_{-	0.05	}	$&$	52.62	

^{+	0.67	}_{-	0.05	}$	&	050915A	&	2.5273	&	21.39	$\pm$	0.59	&$	52.26	^{+	

0.52	}_{-	0.12	}	$&$	52.19	^{+	0.52	}_{-	0.12	}$	\\
090205	&	4.6497	&	10.68	$\pm$	0.69	&$	52.09	^{+	0.59	}_{-	0.09	}	$&$	52.07	

^{+	0.59	}_{-	0.09	}$	&	050908	&	3.35	&	10.80	$\pm$	0.64	&$	52.11	^{+	

0.26	}_{-	0.09	}	$&$	52.06	^{+	0.26	}_{-	0.09	}$	\\
090113	&	1.7493	&	8.80	$\pm$	0.13	&$	52.01	^{+	0.48	}_{-	0.08	}	$&$	51.92	

^{+	0.48	}_{-	0.08	}$	&	050904	&	6.29	&	197.20	$\pm$	2.26	&$	54.13	^{+	

0.22	}_{-	0.13	}	$&$	54.15	^{+	0.22	}_{-	0.13	}$	\\
090102	&	1.547	&	30.69	$\pm$	1.21	&$	53.15	^{+	0.31	}_{-	0.17	}	$&$	53.06	

^{+	0.31	}_{-	0.17	}$	&	050826	&	0.297	&	34.44	$\pm$	1.87	&$	50.53	^{+	

0.52	}_{-	0.24	}	$&$	50.48	^{+	0.52	}_{-	0.24	}$	\\
081228	&	$3.4^{a}$	&	$3.00^{b}$	  	  	&$	52.57	^{+	0.19	}_{-	0.15	}	$$^{c}$&$	52.52	

^{+	0.19	}_{-	0.15	}$	&	050824	&	0.83	&	37.95	$\pm$	4.02	&$	51.19	^{+	

2.47	}_{-	0.12	}	$&$	51.10	^{+	2.47	}_{-	0.12	}$	\\
081222	&	2.77	&	33.48	$\pm$	1.44	&$	53.18	^{+	0.10	}_{-	0.05	}	$&$	53.12	

^{+	0.10	}_{-	0.05	}$	&	050822	&	1.434	&	104.88	$\pm$	2.63	&$	52.37	^{+	

0.64	}_{-	0.03	}	$&$	52.28	^{+	0.64	}_{-	0.03	}$	\\
081221	&	2.26	&	34.23	$\pm$	0.64	&$	53.53	^{+	0.04	}_{-	0.03	}	$&$	53.45	

^{+	0.04	}_{-	0.03	}$	&	050820A	&	2.6147	&	239.68	$\pm$	0.37	&$	53.40	^{+	

0.34	}_{-	0.20	}	$&$	53.33	^{+	0.34	}_{-	0.20	}$	\\
081203A	&	2.1	&	254.28	$\pm$	26.94	&$	53.24	^{+	0.34	}_{-	0.10	}	$&$	53.16	

^{+	0.34	}_{-	0.10	}$	&	050819	&	2.5043	&	46.80	$\pm$	4.85	&$	52.00	^{+	

0.92	}_{-	0.11	}	$&$	51.93	^{+	0.92	}_{-	0.11	}$	\\
081121	&	2.512	&	19.38	$\pm$	0.96	&$	53.21	^{+	0.40	}_{-	0.11	}	$&$	53.14	

^{+	0.40	}_{-	0.11	}$	&	050814	&	5.3	&	27.54	$\pm$	1.71	&$	52.73	^{+	

0.21	}_{-	0.09	}	$&$	52.72	^{+	0.21	}_{-	0.09	}$	\\
081118	&	2.58	&	66.55	$\pm$	5.08	&$	52.46	^{+	0.68	}_{-	0.06	}	$&$	52.39	

^{+	0.68	}_{-	0.06	}$	&	050803	&	0.422	&	88.20	$\pm$	1.35	&$	51.40	^{+	

0.44	}_{-	0.15	}	$&$	51.33	^{+	0.44	}_{-	0.15	}$	\\
081109	&	$0.98^{k}$	&	$221.00^{b}$	  	  	&$	52.61	^{+	0.28	}_{-	0.23	}	$$^{c}$&$	52.52	

^{+	0.28	}_{-	0.23	}$	&	050802	&	1.71	&	14.25	$\pm$	0.60	&$	52.27	^{+	

0.35	}_{-	0.08	}	$&$	52.18	^{+	0.35	}_{-	0.08	}$	\\
081029	&	3.8479	&	169.10	$\pm$	8.55	&$	53.17	^{+	0.25	}_{-	0.20	}	$&$	53.14	

^{+	0.25	}_{-	0.20	}$	&	050801	&	1.56	&	5.88	$\pm$	0.20	&$	51.31	^{+	

0.63	}_{-	0.06	}	$&$	51.22	^{+	0.63	}_{-	0.06	}$	\\
081028	&	3.038	&	275.59	$\pm$	9.68	&$	53.07	^{+	0.12	}_{-	0.08	}	$&$	53.01	

^{+	0.12	}_{-	0.08	}$	&	050730	&	3.969	&	60.48	$\pm$	2.26	&$	52.92	^{+	

0.42	}_{-	0.12	}	$&$	52.88	^{+	0.42	}_{-	0.12	}$	\\
081008	&	1.9685	&	199.32	$\pm$	11.52	&$	52.82	^{+	0.21	}_{-	0.08	}	$&$	52.74	

^{+	0.21	}_{-	0.08	}$	&	050724	&	0.258	&	2.50	$\pm$	0.04	&$	49.96	^{+	

0.49	}_{-	0.08	}	$&$	49.92	^{+	0.49	}_{-	0.08	}$	\\
081007	&	0.5295	&	5.55	$\pm$	0.26	&$	50.87	^{+	0.28	}_{-	0.09	}	$&$	50.79	

^{+	0.28	}_{-	0.09	}$	&	050713A	&	$3.6^{g}$	&	$94.90^{b}$	  	  	&$	54.19	^{+	

0.37	}_{-	0.13	}	$$^{c}$&$	54.15	^{+	0.37	}_{-	0.13	}$	\\
080928	&	1.692	&	284.90	$\pm$	12.16	&$	52.46	^{+	0.38	}_{-	0.08	}	$&$	52.37	

^{+	0.38	}_{-	0.08	}$	&	050607	&	$4^{g}$	&	$48.00^{b}$	  	  	&$	53.09	^{+	

0.38	}_{-	0.05	}	$$^{c}$&$	53.06	^{+	0.38	}_{-	0.05	}$	\\
080916A	&	0.689	&	62.53	$\pm$	3.24	&$	51.92	^{+	0.11	}_{-	0.05	}	$&$	51.84	

^{+	0.11	}_{-	0.05	}$	&	050603	&	2.821	&	9.80	$\pm$	0.39	&$	53.63	^{+	

0.40	}_{-	0.15	}	$&$	53.56	^{+	0.40	}_{-	0.15	}$	\\
080913	&	6.7	&	8.19	$\pm$	0.26	&$	52.85	^{+	0.41	}_{-	0.09	}	$&$	52.87	

^{+	0.41	}_{-	0.09	}$	&	050525	&	0.606	&	9.10	$\pm$	0.04	&$	52.32	^{+	

0.02	}_{-	0.02	}	$&$	52.24	^{+	0.02	}_{-	0.02	}$	\\
080905B	&	2.374	&	103.97	$\pm$	4.68	&$	52.55	^{+	0.39	}_{-	0.08	}	$&$	52.47	

^{+	0.39	}_{-	0.08	}$	&	050505	&	4.27	&	60.20	$\pm$	1.35	&$	53.21	^{+	

0.38	}_{-	0.10	}	$&$	53.18	^{+	0.38	}_{-	0.10	}$	\\
080810	&	3.35	&	453.15	$\pm$	5.09	&$	53.56	^{+	0.27	}_{-	0.19	}	$&$	53.50	

^{+	0.27	}_{-	0.19	}$	&	050502B	&	$5.2^{i}$	&	$16.62^{e}$		  	&$	52.82	^{+	

0.04	}_{-	0.04	}	$$^{f}$&$	52.81	^{+	0.04	}_{-	0.04	}$	\\
080805	&	1.505	&	111.84	$\pm$	9.11	&$	52.62	^{+	0.22	}_{-	0.17	}	$&$	52.53	

^{+	0.22	}_{-	0.17	}$	&	050416A	&	0.6535	&	2.91	$\pm$	0.18	&$	51.00	^{+	

0.19	}_{-	0.09	}	$&$	50.92	^{+	0.19	}_{-	0.09	}$	\\
080804	&	2.2	&	61.74	$\pm$	8.81	&$	53.21	^{+	0.45	}_{-	0.18	}	$&$	53.13	

^{+	0.45	}_{-	0.18	}$	&	050412	&	$4.5^{g}$	&	$26.50^{b}$	  	  	&$	54.00	^{+	

0.79	}_{-	0.26	}	$$^{c}$&$	53.98	^{+	0.79	}_{-	0.26	}$	\\
080721	&	2.602	&	29.92	$\pm$	2.29	&$	54.06	^{+	0.42	}_{-	0.20	}	$&$	53.99	

^{+	0.42	}_{-	0.20	}$	&	050406	&	$2.7^{i}$	&	$4.79^{e}$		  	&$	51.56	^{+	

0.09	}_{-	0.09	}	$$^{f}$&$	51.49	^{+	0.09	}_{-	0.09	}$	\\
080710	&	0.845	&	139.05	$\pm$	10.01	&$	51.91	^{+	0.46	}_{-	0.23	}	$&$	51.82	

^{+	0.46	}_{-	0.23	}$	&	050401	&	2.9	&	34.41	$\pm$	0.34	&$	53.52	^{+	

0.35	}_{-	0.09	}	$&$	53.46	^{+	0.35	}_{-	0.09	}$	\\
080707	&	1.23	&	30.25	$\pm$	0.43	&$	51.55	^{+	0.52	}_{-	0.07	}	$&$	51.45	

^{+	0.52	}_{-	0.07	}$	&	050319	&	3.24	&	153.55	$\pm$	2.20	&$	52.67	^{+	

0.62	}_{-	0.05	}	$&$	52.62	^{+	0.62	}_{-	0.05	}$	\\
080607	&	3.036	&	83.66	$\pm$	0.83	&$	54.46	^{+	0.20	}_{-	0.14	}	$&$	54.40	

^{+	0.20	}_{-	0.14	}$	&	050318	&	1.44	&	30.96	$\pm$	0.09	&$	52.08	^{+	

0.08	}_{-	0.09	}	$&$	51.98	^{+	0.08	}_{-	0.09	}$	\\
080605	&	1.6398	&	19.57	$\pm$	0.32	&$	53.33	^{+	0.19	}_{-	0.08	}	$&$	53.24	

^{+	0.19	}_{-	0.08	}$	&	050315	&	1.949	&	94.60	$\pm$	1.66	&$	52.77	^{+	

0.48	}_{-	0.01	}	$&$	52.68	^{+	0.48	}_{-	0.01	}$	\\
080604	&	1.416	&	125.28	$\pm$	5.37	&$	51.86	^{+	0.46	}_{-	0.09	}	$&$	51.77	

^{+	0.46	}_{-	0.09	}$	&	050223	&	0.5915	&	17.38	$\pm$	0.60	&$	50.87	^{+	

0.29	}_{-	0.08	}	$&$	50.79	^{+	0.29	}_{-	0.08	}$	\\
080603B	&	2.69	&	59.50	$\pm$	0.51	&$	52.80	^{+	0.07	}_{-	0.07	}	$&$	52.74	

^{+	0.07	}_{-	0.07	}$	&	050126	&	1.29	&	28.71	$\pm$	1.91	&$	51.90	^{+	

0.58	}_{-	0.12	}	$&$	51.81	^{+	0.58	}_{-	0.12	}$	\\
\hline
\end{longtable}
\footnotesize
$^{a}$Redshift from Greiner et al. (2011).
$^{b}$$T_{90}$ taken from Robertson \& Ellis (2012).
$^{c}$$E_{\rm iso}$ taken from Robertson \& Ellis (2012).
$^{d}$Redshift from Perley \& Perley (2013).
$^{e}$$T_{90}$ taken from Sakamoto et al. (2011).
$^{f}$$E_{\rm iso}$ calculated from the fluence provided by Sakamoto et al. (2011).
$^{g}$Dark GRB redshift limit from Perley et al. (2009).
$^{h}$Redshift from Perley et al. (2009).
$^{i}$Redshift from Hjorth et al. (2012).
$^{j}$Dark GRB redshift limit from Greiner et al. (2011).
$^{k}$Redshift from Kr$\ddot{\rm u}$hler et al. (2011).
\end{small}
\end{center}

Since we will use the cumulative redshift distribution $N(<z)$ of this sample
as the basis for our analysis, it is important to consider its uncertainties.
Redshift measurements are strongly biased towards optically bright afterglows,
and are more easily made when the afterglow is not obscured by dust (see, e.g., Greiner
et al. 2011). The phenomenon of so-called dark GRBs with suppressed optical
counterparts could influence whether the observed $N(< z)$ is representative
of that for all long-duration GRBs. Perley et al. (2009) have considered this
important issue by attempting to constrain the redshift distribution of dark
GRBs through deep searches that successfully located faint optical and near-infrared
counterparts. The Perley et al. (2009) work provides us with one redshift and
nine redshift upper limits for a subsample of dark GRBs in our catalog. Greiner
et al. (2011) and Kr$\ddot{\rm u}$hler et al. (2011) have pursued this effort in parallel
and have provided three additional redshifts and one redshift upper limit for
dark GRBs in our catalog. Via host galaxy measurements, Hjorth et al. (2012) and
Perley \& Perley (2013) have also provided nine additional redshifts for dark GRBs
that we have added to our catalog. We assume that the subsamples of dark GRBs with redshift upper
limits presented by Perley et al. (2009), Greiner et al. (2011), and Kr$\ddot{\rm u}$hler
et al. (2011) are representative of that class, and therefore optionally incorporate
those limits to characterize the effects of possible incompleteness of the Swift
sample with firm redshift determinations.

\begin{figure*}
\hskip0.3in\includegraphics[angle=0,scale=0.5]{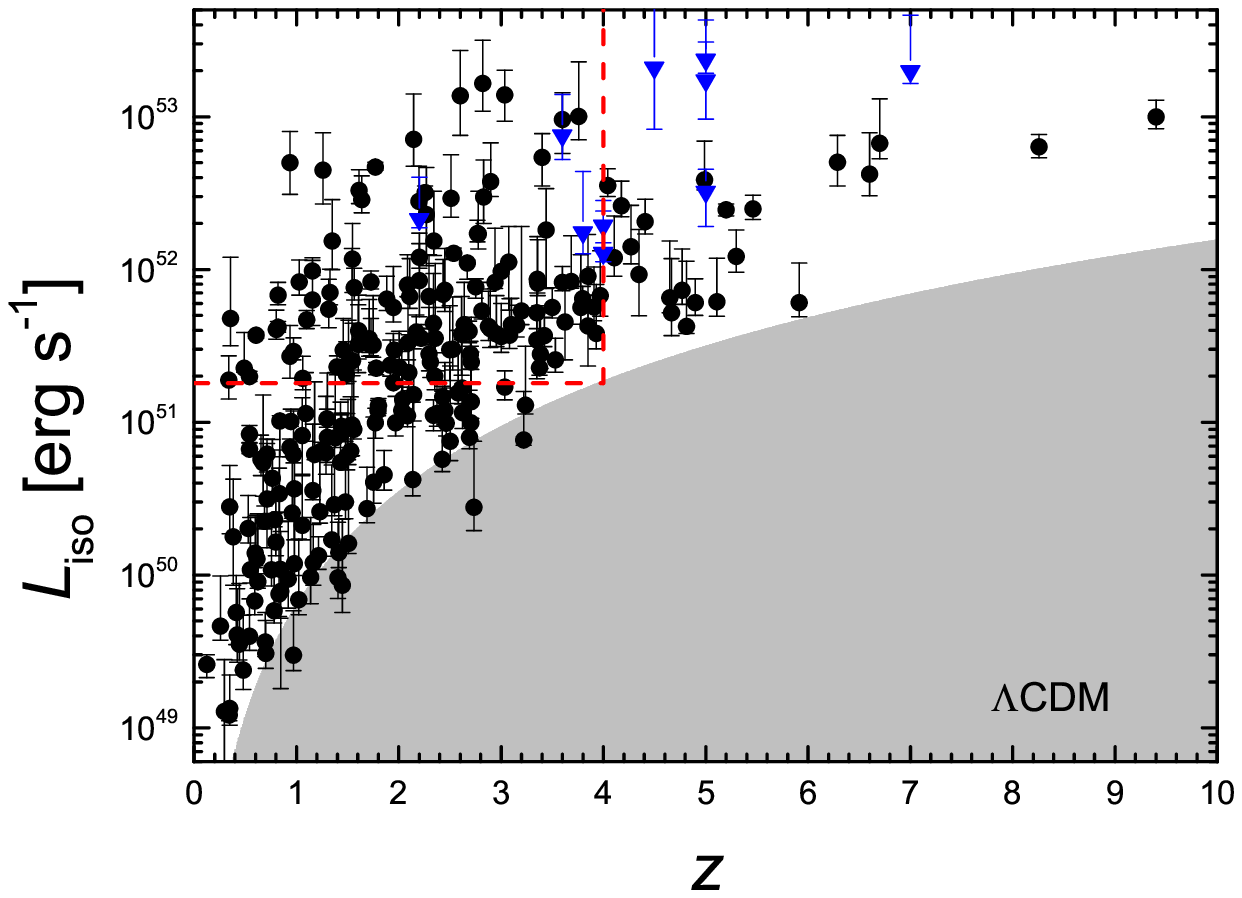}
\includegraphics[angle=0,scale=0.5]{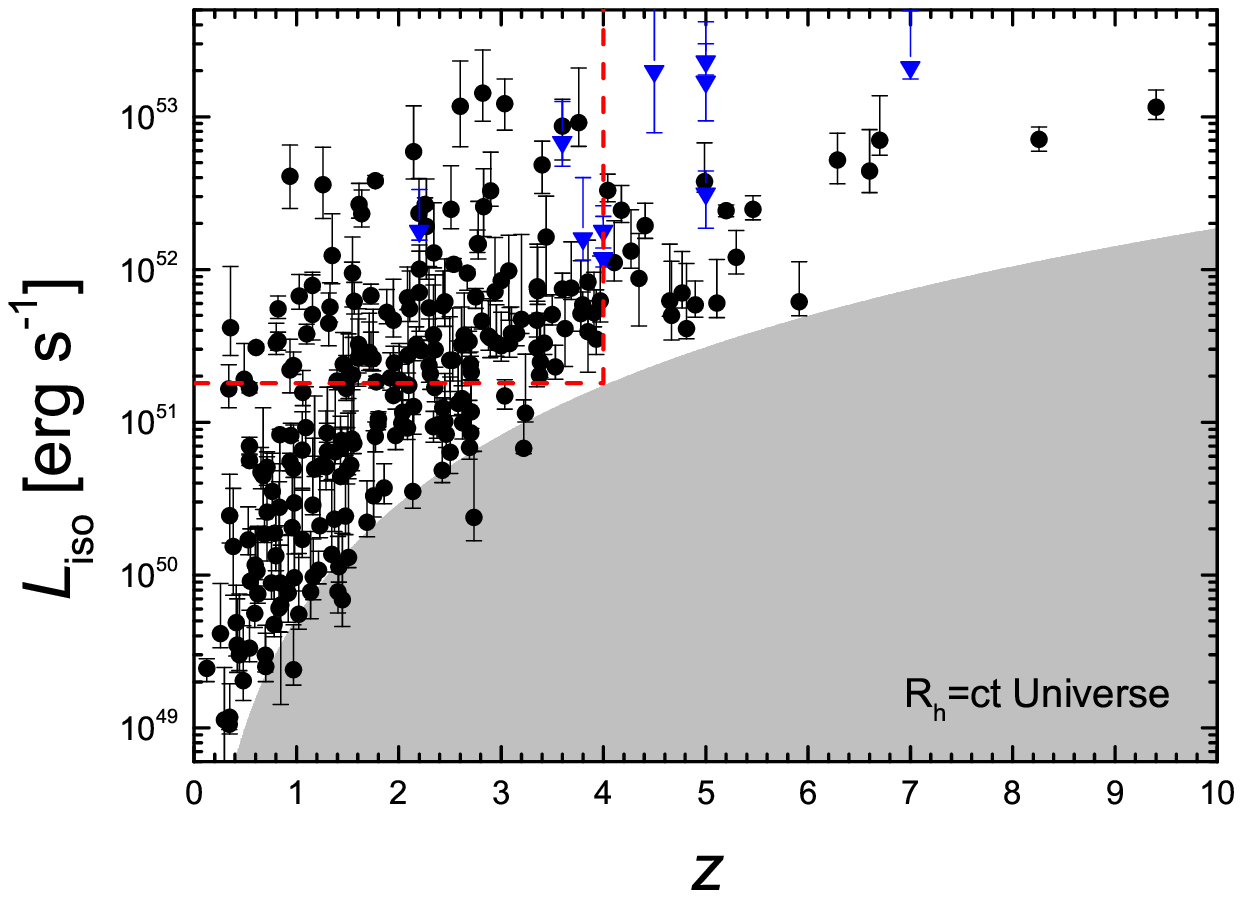}
\caption{The luminosity-redshift distribution of 254 \emph{Swift} GRBs
in $\Lambda$CDM (left panel) and $R_{\rm h}=ct$ (right panel). The blue dots
represent the bursts with redshift upper limits. The shaded regions
represent the luminosity threshold adopted in our calculations
(see text and equation~\ref{Llimit}).}\label{LZ}
\end{figure*}

Our final sample includes 254 GRBs, whose luminosity-redshift distribution
is shown in Figure~\ref{LZ}. A determination of $E_{\rm iso}$ requires the
assumption of a particular cosmological model. In this figure, we show the
resulting distributions for both $\Lambda$CDM (left panel) and
$R_{\rm h}=ct$ (right panel). As presented in the various sources used
to compile our catalog, quantities such as $E_{\rm iso}$ are estimated
assuming a $\Lambda$CDM cosmology.  Here, we must therefore recalibrate
them for use in $R_{\rm h}=ct$. The differences between these two
models\footnote{See also Melia (2012b) for a more pedagogical description
of the $R_{\rm h}=ct$ Universe.} are summarized in Melia (2012a,2013a,2013b),
Melia \& Shevchuk (2012), Melia \& Maier (2013), and Wei et al.
(2013). The luminosity distance in $\Lambda$CDM is given by the expression
\begin{equation}
D_{L}^{\Lambda {\rm CDM}}(z)={c\over H_{0}}{(1+z)\over\sqrt{\mid\Omega_{k}\mid}}\; sinn\left\{\mid\Omega_{k}\mid^{1/2}
\times\int_{0}^{z}{dz\over\sqrt{(1+z)^{2}(1+\Omega_{m}z)-z(2+z)\Omega_{\Lambda}}}\right\}\;,
\end{equation}
where $c$ is the speed of light, and $H_{0}$ is the Hubble constant at the present time.
In this equation, $\Omega_m\equiv\rho_m/\rho_c$ is the energy density of
matter written in terms of today's critical density, $\rho_c\equiv 3c^2 H_0^2/8\pi G$.
Also, $\Omega_{\Lambda}$ is the similarly defined density of dark energy, and
$\Omega_{k}$ represents the spatial curvature of the Universe---appearing as a
term proportional to the spatial curvature constant $k$ in the Friedmann equation.
In addition, $sinn$ is $\sinh$ when $\Omega_{k}>0$ and $\sin$  when $\Omega_{k}<0$.
For a flat Universe with $\Omega_{k}=0$, this equation simplifies to the form
$(1+z)c/H_{0}$ times the integral.

In the $R_{\rm h}=ct$ Universe, the luminosity
distance is given by the much simpler expression
\begin{equation}
D_{L}^{R_{\rm h}=ct}=\frac{c}{H_{0}}(1+z)\ln(1+z)\;.
\end{equation}
The factor $c/H_0$ is in fact the gravitational horizon $R_{\rm h}(t_0)$ at the
present time, so we may also write the luminosity distance as
\begin{equation}
D_{L}^{R_{\rm h}=ct}=R_{\rm h}(t_0)(1+z)\ln(1+z)\;.
\end{equation}
We have found the equivalent isotropic energy in the $R_{\rm h}=ct$
Universe using the expression
\begin{equation}
{E}_{\rm iso}^{R_{\rm h}=ct}=E_{\rm iso}^{\Lambda{\rm CDM}}
\times \left({D^{R_{\rm h}=ct}_{L}\over D^{\Lambda{\rm CDM}}_{L}}\right)^2\;,
\end{equation}
where $E_{\rm iso}^{\Lambda{\rm CDM}}$ is the  previously published value.

\section{The Model}

The observed rate of GRBs per unit time at redshifts $\in (z,z+dz)$ with
luminosity $\in (L,L+dL)$ is given by
\begin{equation}\label{density}
\frac{dN}{dt\,dz\,dL}=\frac{\dot{\rho}_{\rm GRB}(z)}{1+z}
\frac{\Delta\Omega}{4\pi}\frac{dV_{\rm com}(z)}{dz}\,\Phi(L)\;,
\end{equation}
where $\dot{\rho}_{\rm GRB}(z)$ is the co-moving rate density of
GRBs, $\Phi(L)$ is the beaming-convolved luminosity function (LF), the
factor $(1+z)^{-1}$ accounts for the cosmological time dilation
and $\Delta\Omega=1.4$ sr is the solid angle covered on the sky by
\emph{Swift} (Salvaterra \& Chincarini 2007). The co-moving volume
is calculated using
\begin{equation}\label{volume}
\frac{dV_{\rm com}}{dz}=4\pi D_{\rm com}^{2}\frac{dD_{\rm com}}{dz}.
\end{equation}
In the standard ($\Lambda$CDM) model, the co-moving luminosity
distance is given as
\begin{equation}\label{distanc}
D_{\rm com}^{\rm \Lambda CDM}(z)\equiv\frac{c}{H_{0}}\int_{0}^{z}\frac{dz'}{\sqrt{\Omega_{\rm m}
(1+z')^{3}+\Omega_{\Lambda}}}\;,
\end{equation}
where we now adopt concordance values of the cosmological parameters:
$H_{0}=70$ km $\rm s^{-1}$ $\rm Mpc^{-1}$, $\Omega_{\rm m}=0.3$,
and $\Omega_{\Lambda}=0.7$, and assume a spatially flat expansion.
In the $R_{\rm h}=ct$ Universe, the co-moving luminosity distance is
given by the much simpler expression
\begin{equation}
D_{\rm com}^{R_{\rm h}=ct}(z)=\frac{c}{H_{0}}\ln(1+z)
\end{equation}
which, as we have noted previously, has only one free parameter---the
Hubble constant $H_{0}$. For the sake of consistency, we will adopt the
standard $H_{0}=70$ km $\rm s^{-1}$ $\rm Mpc^{-1}$ throughout this
analysis.

As discussed above, we assume that the GRB rate density is related to
the cosmic SFR density $R_{\rm SF}(z)$ and a possible evolution effect
$f(z)$, given as
\begin{equation}\label{density1}
\dot{\rho}_{\rm GRB}(z)=k_{\rm GRB}R_{\rm SF}(z)f(z)\;,
\end{equation}
where $k_{\rm GRB}$ is the GRB formation efficiency to be determined from the observations.

Because of the faintness of sub-luminous galaxies at high redshifts,
as well as the uncertainty of the dust extinction (in terms of the
amount of dust as well as the dust attenuation law), it is difficult
to observe LBG's at high redshifts. Consequently, the LBG samples are
incomplete, and the star formation history at $z \ga 4$ is not well
constrained by the data. For relatively low
redshifts ($z \la 4$), the star formation rate density $R_{\rm SF}$
has been fitted with a piecewise power law (Hopkins
\& Beacom 2006; Li 2008), which in $\Lambda$CDM (with the concordance,
WMAP parameters) may be written
\begin{equation}\label{SFRone}
\log_{10} R_{\rm SF}^{\rm \Lambda CDM}(z)=a+b\log_{10}(1+z)\;,
\end{equation}
where
\begin{equation}
  (a,b) = \left\lbrace \begin{array}{ll}(-1.70, 3.30), ~~~~~~~~~~~z <0.993\\
                                           (-0.727, 0.0549), ~~~~~0.993< z <3.8, \\
\end{array} \right.
\end{equation}
and $R_{\rm SF}$ is in units of $\rm M_{\bigodot}$ $\rm yr^{-1}$ $\rm Mpc^{-3}$.
To convert from one cosmology to another, our procedure is as follows: the co-moving
volume is proportional to the co-moving distance cubed, $V_{\rm com} \propto
D_{\rm com}^{3}$, and the co-moving volume between redshifts $z-\Delta z$ and
$z+\Delta z$ is $V_{\rm com}(z, \Delta z) \propto D_{\rm com}^{3}(z+\Delta z)
-D_{\rm com}^{3}(z-\Delta z)$. Since the luminosity is proportional to the
co-moving distance squared, $L \propto D_{\rm com}^{2}$, the SFR density for
a given redshift range is (Hopkins 2004)
\begin{equation}
R_{\rm SF}(z)\propto\frac{L(z)}{V_{\rm com}(z, \Delta z)}\propto\frac{D_{\rm com}^{2}(z)}
{D_{\rm com}^{3}(z+\Delta z)-D_{\rm com}^{3}(z-\Delta z)}.
\end{equation}
Thus, the SFR in the redshift range $z =0 - 3.8$ for the $R_{\rm h}=ct$ Universe becomes
\begin{equation}\label{SFRtwo}
\log R_{\rm SF}^{R_{\rm h}=ct}(z)=a+b\log(1+z)\;,
\end{equation}
where
\begin{equation}
  (a,b) = \left\lbrace \begin{array}{ll}(-1.70, 3.52), ~~~~~~~~~~~z <0.993\\
                                           (-0.507, -0.46), ~~~~~0.993< z <3.8. \\
\end{array} \right.
\end{equation}

For the GRB luminosity function (LF) $\Phi(L)$, several models have been
adopted in the literature: a single power law with an exponential cut-off at low luminosity
(exponential LF), a broken power law, and a Schechter function. Here, we use the exponential LF
\begin{equation}\label{Schechter}
\Phi(L)\propto\left(\frac{L}{L_{\star}}\right)^{-a_{\rm L}}\exp\left(-\frac{L_{\star}}{L}\right)\;,
\end{equation}
where $a_{\rm L}$ is the power-law index and $L_{\star}$ is the cutoff luminosity.
The normalization constant of the LF is calculated assuming a minimum luminosity
$L_{\rm min}=10^{49}$ erg s$^{-1}$. The LF will be taken to be non-evolving in this paper.

Finally, when considering an instrument having a flux threshold,
the expected number of GRBs with luminosity $L_{\rm iso}>L_{\rm lim}$
and redshift $z\in (z_{1},z_{2})$ during an observational period {\em T}
should be
\begin{equation}\label{Number}
\begin{split}
N=\frac{\Delta\Omega\;
T}{4\pi}
\int_{z_{1}}^{z_{2}}\frac{\dot{\rho}_{\rm GRB}(z)}{1+z}\frac{dV_{\rm com}(z)}{dz}\;dz\int_{\rm max[L_{\rm min}, L_{\rm lim}(z)]}^{\infty}\Phi(L)\;dL\;.
\end{split}
\end{equation}
The luminosity threshold appearing in equation~(\ref{Number}) may be
approximated using a bolometric energy flux limit $F_{\rm lim}=1.2\times10^{-8}$
erg $\rm cm^{-2}$ $\rm s^{-1}$ (Li 2008), for which
\begin{equation}\label{Llimit}
L_{\rm lim}=4\pi D_{\rm L}^{2} F_{\rm lim}\;,
\end{equation}
where $D_{\rm L}$ is the luminosity distance to the burst (either $D_{\rm L}^{\Lambda{\rm CDM}}$
or $D_{\rm L}^{R_{\rm h}=ct}$, as the case may be).

\section{A Comparative Analysis of $\Lambda$CDM and The $R_{\rm h}=ct$ Universe}

\subsection{A possible evolutionary effect}
The \emph{Swift}/BAT trigger is quite complex. Its algorithm
has two modes: the count rate trigger and the image trigger
(Fenimore et al. 2003; Sakamoto et al. 2008, 2011). Rate triggers
are measured on different timescales (4 ms to 64 s), with a single
or several backgrounds. Image triggers are found by summing images
over various timescales and searching for uncataloged sources. So
the sensitivity of the BAT is very difficult to parametrize exactly
(Band 2006). Moreover, although the rate density $R_{\rm SF}(z)$ is
now reasonably well measured from $z=0$ to $4$, it is not well
constrained at $z \ga 4$. To avoid the complications that would
arise from the use of a detailed treatment of the \emph{Swift}
threshold and the star formation rate at high-\emph{z}, we will
adopt a model-independent approach by selecting only GRBs with
$L_{\rm iso} \geq L_{\rm lim}$ and $z<4$, as Kistler et al. (2008)
did in their treatment. The cut in luminosity\footnote{Note
that although the luminosity distances are formulated differently
in the two cosmologies we are examining here, distance measures
in the optimized $\Lambda$CDM model are very close to those
in $R_{\rm h}=ct$, so this cutoff does not bias either model.}
is chosen to be equal
to the threshold at the highest redshift of the sample, i.e.,
$L_{\rm lim}\approx1.8\times10^{51}$ erg $\rm s^{-1}$.
The cuts in luminosity and redshift minimize selection effects
in the GRB data. With these conditions, our final tally of GRBs is 118 for
$\Lambda$CDM and 111 in $R_{\rm h}=ct$. These data
are delimited by the red dashed lines in Figure~2.

\begin{figure*}
\includegraphics[angle=0,scale=0.50]{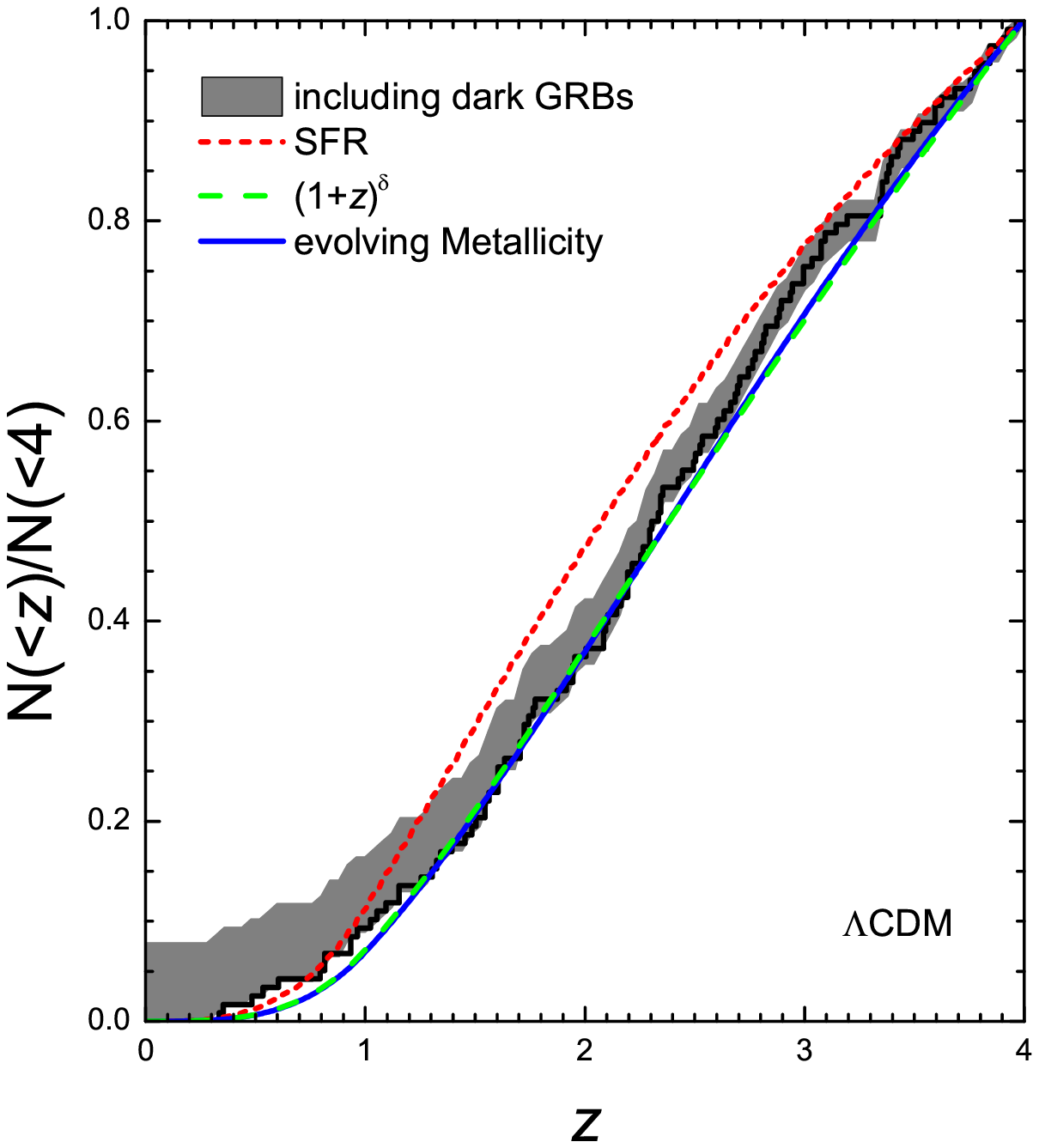}\hskip 0.3in
\includegraphics[angle=0,scale=0.50]{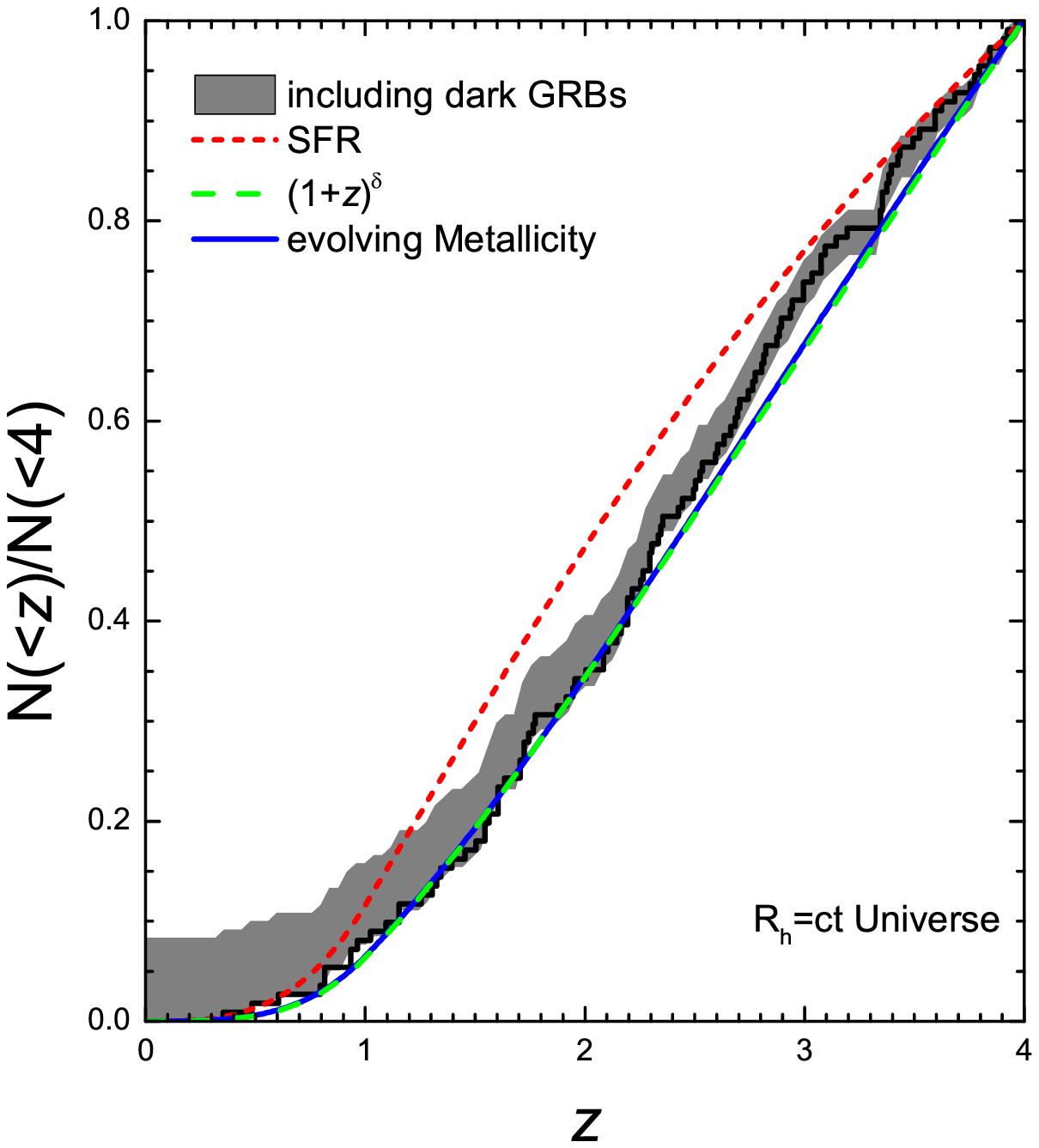}
    \caption{Left: Cumulative distribution of 118 \emph{Swift} GRBs with $z<4$
    and $L_{\rm iso}>1.8\times10^{51}$ erg $s^{-1}$, assuming the
    standard ($\Lambda$CDM) model. The black steps and the gray area indicate
    the cumulative distribution of GRBs with firm reshifts and the uncertainty
    owing to dark GRBs. Three fitting results using equation~(\ref{SMALL})
    are also shown: the red dotted line corresponds to the SFR on its own (i.e., $f(z)$ is
    constant), the pink dashed line corresponds to $f(z) \propto (1+z)^\delta$, with
    $\delta=0.80$, and the blue solid line corresponds to $f(z) \propto \Theta(\epsilon,z)$,
    with $\epsilon=0.52$. Right: Same as the left panel, but for 111 \emph{Swift} GRBs
    and with $\delta=1.03$ and $\epsilon=0.44$ in the $R_{h}=ct$ Universe.}\label{small4}
\end{figure*}

Now, since $L_{\rm lim}$ is constant, the integral of the LF
in equation~(\ref{Number}) can be treated as a constant coefficient,
no matter what the specific form of $\Phi(L)$ happens to be.
That is, we may write
\begin{equation}\label{SMALL}
\begin{split}
N(<z)\propto \int_{0}^{z}R_{\rm SF}(z)\frac{f(z)}{1+z}\frac{dV_{\rm com}}{dz}\;dz\;.
\end{split}
\end{equation}
Figure~\ref{small4} shows the cumulative redshift distribution of
observed GRBs (steps), normalized over the redshift range
$0<z<4$. The gray-shaded region shows how the distribution
shifts in the limiting cases of all dark GRBs occurring
at $z=0$ or the upper redshift limits determined by Perley et al.
(2009), Greiner et al. (2011), and Kr$\ddot{\rm u}$hler et al. (2011).
In the left panel of this figure, we assume the
$\Lambda$CDM cosmology, and compare the observed GRB
cumulative redshift distribution with three types of redshift evolution,
characterized through the function $f(z)$. If this function is constant
(dotted red line), the expectation from the SFR alone (i.e., the non-evolution case)
is incompatible with the observations. If we parameterize the possible
evolution effect as $f(z)\propto(1+z)^{\delta}$, we find that the $\chi^{2}$
statistic is minimized for $\delta=0.8$, which is consistent with that
of Robertson \& Ellis (2012) and Wang (2013). The weak redshift
evolution ($\delta=0.8$) can reproduce the observed cumulative
GRB rate density best (dashed green line). At the $2\sigma$ confidence
level, the value of $\delta$ lies in the range $0.07<\delta<1.53$.
In the limiting case where all the dark GRBs are local, the power-law
index is constrained to be $-0.56<\delta<0.98$ at the $2\sigma$
confidence level. The peak probability occurs for $\delta=0.21$.
Instead, if all the dark GRBs are at their maximum possible redshift,
the power-law index moves to $0.19<\delta<1.67$ ($2\sigma$)
with a peak near $\delta\approx0.93$. Clearly, the additional uncertainty
arising from the inclusion of dark GRBs is an important consideration.
If dark GRBs occur at their maximum allowed redshifts, the distribution
is more heavily weighted towards higher values of $z$ and would therefore
indicate a stronger redshift dependence of the relationship between
the GRB rate and the SFR. We will discuss the third type of evolution shortly.

The collapsar model predicts that
long bursts should occur preferentially in metal poor environments. From
a theoretical standpoint, this is not surprising since lower metallicity
leads to weaker stellar winds and hence less angular momentum loss,
resulting in the retention of rapidly rotating cores in stars at the time
of their explosion, as implied by simulations of the collapsar model for GRBs
(e.g., Woosley 1993b; MacFayden \& Woosley 1999; Yoon \& Langer 2005).
It has therefore been suggested that the observationally required evolution
may be due mainly to the cosmic evolution in metallicity.

According to Langer \& Norman (2006), the fractional mass density belonging
to metallicity below $Z=\epsilon Z_{\odot}$ (where $Z_{\odot}$ is the solar
metal abundance, and $\epsilon$ is determined by the metallicity threshold
for the production of GRBs) at a given redshift \emph{z} can be calculated
using $\Theta(\epsilon,z)=\hat{\Gamma}(\kappa+2,
\epsilon^{\beta} 10^{0.15 \beta z})/\Gamma(\kappa+2)$,
where $\kappa=-1.16$ is the power-law index in the Schechter distribution
function of galaxy stellar masses (Panter et al. 2004), $\beta=2$ is the
slope of the linear bisector fit to the galaxy stellar mass-metallicity
relation (Savaglio 2006), and $\hat{\Gamma}(a,x)$ and $\Gamma(x)$ are the
incomplete and complete Gamma functions, respectively. To test this interpretation
of the anomalous evolution, we parameterize the evolution function as
$f(z) \propto \Theta(\epsilon,z)$, and show the result of an evolving
metallicity as a blue line in the left panel of Figure~\ref{small4}.
This theoretical curve agrees very well with the observations. The best
fit to the observations yields $\epsilon=0.52$. At the $2\sigma$ confidence
level, the value of $\epsilon$ lies in the range $0.19<\epsilon<0.85$. A comparison between
this curve and that obtained with $f(z)=(1+z)^{0.8}$ shows that
the differences between these two fits is not very significant. Therefore,
we confirm that the anomalous evolution in $\Lambda$CDM may be due
to an evolving metallicity. However, in contrast to previous studies
that suggest a metallicity cut of $Z_{\rm th} \la 0.3Z_{\odot}$
(Woosley \& Heger 2006; Langer \& Norman 2006; Salvaterra \& Chincarini
2007; Li 2008; Campisi et al. 2010; Salvaterra et al. 2012), we find that only the higher
metallicity cut $Z_{\rm th}=0.52Z_{\odot}$ is consistent with
the data, in agreement with the conclusions of Hao \& Yuan (2013). It is worth mentioning
that the higher metallicity cut is also more consistent with recent studies of the
long GRB host galaxies (Graham et al. 2009; Levesque et al. 2010a,b; Michalowski
et al. 2012).

The right panel of Figure~\ref{small4} shows the cumulative redshift distribution
of 111 \emph{Swift} GRBs with $L_{\rm lim}=1.8\times10^{51}$ erg $\rm
s^{-1}$ and $z<4$ in the $R_{\rm h}=ct$ Universe. The result of our fitting
from the SFR alone (i.e., with a constant $f[z]$) is shown as a dotted red
line, which again is incompatible with the observations. An additional
evolutionary effect, parametrized as $f(z)\propto(1+z)^{1.03}$ is required
(dashed green line). At the $2\sigma$ confidence level, the value of $\delta$
lies in the range $0.32<\delta<1.71$. If the dark GRB sample with redshift limits
is assumed to be local ($z\approx0$), the $2\sigma$ interval is $-0.38<\delta<1.12$
with a peak near $\delta=0.37$. Instead, if all dark GRBs are at their maximum
possible redshift, the power-law index moves to $0.43<\delta<1.87$ ($2\sigma$)
with a peak near $\delta\approx1.15$. Clearly, if dark GRBs occur at their
maximum allowed redshifts, the distribution is more heavily weighted toward
higher redshifts and the extra redshift evolution effect still exists in
the $R_{\rm h}=ct$ Universe. If we instead designate the evolutionary effect as $f(z) \propto
\Theta(\epsilon,z)$, the evolving metallicity agrees very well with
the observations (blue line). The best fit to the observations
yields $\epsilon=0.44\pm0.28(2\sigma)$. Clearly, the evolutionary effect in both the
$\Lambda$CDM and the $R_{\rm h}=ct$ cosmologies can be
accounted for with a metallicity cutoff at $Z_{\rm th}$ ($0.52Z_{\odot}$
for the former and $0.44Z_{\odot}$ for the latter).

In the next section, we will consider the implications of these
findings for the star-formation history, assuming that GRBs trace
both star formation and a possible evolutionary effect.
We will adopt the best fitting values $\delta=0.80$ or $\epsilon=0.52$ for a reasonable
description of the evolutionary effect in $\Lambda$CDM, and $\delta=1.03$ or $\epsilon=0.44$
in the $R_{\rm h}=ct$ Universe.

\subsection{Constraints on the high-\emph{z} star formation history in $\Lambda$CDM and the $R_{\rm h}=ct$ Universe}
The SFR is well measured at low-\emph{z} now. For high-\emph{z} ($z \ga 4$), a
decrease to the SFR was seemingly implied by the work of Hopkins \& Beacom (2006),
which was confirmed by observations of LBGs and GRBs. Nonetheless,
given the poor coverage of these remote regions, the SFR trends towards
high-\emph{z} are still rather ambiguous. For this reason, previous
studies have included all possibilities: one in which the star-formation
history continues to plateau, or in which it drops off, or even increases with
increasing redshift (see, e.g., Daigne et al. 2006). In our analysis, we will
introduce a free parameter $\alpha$ to parameterize the high-\emph{z} history
as a power law at redshifts $z\ge 3.8$:
\begin{equation}\label{HSFR}
  R_{\rm SF}(z) = \left\lbrace \begin{array}{ll}0.20\left(\frac{1+z}{4.8}\right)^{\alpha}, ~~~~~{\rm for~~~\Lambda CDM}\\
                                                0.15\left(\frac{1+z}{4.8}\right)^{\alpha}, ~~~~~{\rm for~~~R_{\rm h}=ct}, \\
\end{array} \right.
\end{equation}
and we will attempt to constrain this index $\alpha$ using the GRB observations.
The normalization constant in this expression is set by the requirement that
$R_{\rm SF}$ be continuous across $z=3.8$.

We optimize the values of each model's free parameters, including the index
$\alpha$ of high-\emph{z} SFR, the GRB formation efficiency $k_{\rm GRB}$, and the
GRB LF, by minimizing the $\chi^{2}$ statistic jointly fitting the observed redshift
distribution and luminosity distribution of bursts in our sample with firm
measurements of their redshift. The observed number of GRBs in each
redshift bin $z\in(z_{1},z_{2})$ is given by equation~(17), while, the
observed number of events in each luminosity bin
$L_{\rm iso}\in(L_{\rm 1},L_{\rm 2})$ is given by

\begin{equation}\label{LNumber}
\begin{split}
N_{(L_{1}, L_{2})}=\frac{\Delta\Omega\; T}{4\pi}\int_{L_{1}}^{L_{2}}\Phi(L)\;dL
\int_{0}^{z_{\rm max}(L)}\frac{\dot{\rho}_{\rm GRB}(z)}{1+z}\frac{dV_{\rm com}(z)}{dz}\,dz\;,
\end{split}
\end{equation}
where $T\sim8.6$ yr is the observational period, and $z_{\rm max}(L_{\rm iso})$
is the maximum redshift out to which a GRB of luminosity $L_{\rm iso}$ can be
detected; this is obtained by solving the equation $L_{\rm lim}(z)=L_{\rm iso}$ for
each assumed cosmology.

\begin{center}
\begin{table}
\caption{Best-fitting Results in Different Cosmological Models.}
\label{2}
\begin{tabular}{lcccccc}
\hline
Model&$\alpha$&$k_{\rm GRB}$&$L_{\star}$&$a_{L}$&$\chi^{2}$&AIC\\
 & &($10^{-9}$ $M_{\odot}^{-1}$)&($10^{49}$ erg $\rm s^{-1}$)& & & \\
\hline
  &		&	  $\Lambda$CDM			 \\
  \hline
No evol                            & $-2.48_{-1.46}^{+1.45}$	&	$6.21_{-0.88}^{+0.36}$ 	&	$1.03_{-0.39}^{+0.39}$	&	$1.41_{-0.04}^{+0.03}$	&	66.5	            & 74.5       \\
Density evol ($\delta=0.80$)       & $-3.06_{-2.01}^{+2.01}$	&	$4.39_{-1.12}^{+0.67}$ 	&	$0.46_{-0.48}^{+0.48}$	&	$1.51_{-0.07}^{+0.08}$	&	55.4	            & 63.4       \\
Metallicity evol ($\epsilon=0.52$) & $-2.41_{-2.09}^{+1.87}$	&	$13.3_{-2.6}^{+3.1}$ 	&	$1.19_{-0.29}^{+0.29}$	&	$1.51_{-0.05}^{+0.09}$	&	56.0	            & 64.0       \\
 \hline
  &		&	 $R_{\rm h}=ct$			 \\
 \hline
No evol                            & $-3.27_{-1.39}^{+1.44}$	&	$7.22_{-1.05}^{+0.33}$ 	&	$1.11_{-0.68}^{+0.68}$	&	$1.40_{-0.04}^{+0.03}$	&	67.4	            & 75.4       \\
Density evol ($\delta=1.03$)       & $-4.47_{-2.34}^{+2.30}$	&	$3.77_{-1.01}^{+0.49}$ 	&	$1.06_{-0.66}^{+0.66}$	&	$1.54_{-0.05}^{+0.11}$	&	58.6	            & 66.6       \\
Metallicity evol ($\epsilon=0.44$) & $-3.60_{-2.45}^{+2.45}$	&	$19.5_{-4.4}^{+4.2}$ 	&	$1.11_{-0.26}^{+0.34}$	&	$1.50_{-0.02}^{+0.14}$	&	54.3	            & 62.3       \\
\hline
\end{tabular}
\medskip \\
$^{\rm Notes.}$The total number of data points in the fit is 42, including 33 points for the redshift distribution and 9 points for the luminosity distribution.
\end{table}
\end{center}

We report the best-fit parameters together
with their $1\sigma$ confidence level for different models in Table~2. In the last
two columns, we give the total
$\chi^{2}$ value (i.e., the sum of the $\chi^{2}$ values obtained from the fit
of the redshift and luminosity distributions) and the Akaike information criterion (AIC)
score, respectively. For each fitted model, the AIC is given by ${\rm AIC} = \chi^2 + 2k$,
where $k$ is the number of free parameters. If there are three models
for the data, $\mathcal{M}_1$, $\mathcal{M}_2$, and $\mathcal{M}_3$,
and they have been separately fitted, the one with the least resulting AIC is the
one favored by this criterion. A more quantitative ranking of models can be
computed as follows. If ${\rm AIC}_\alpha$ comes from model~$\mathcal{M}_\alpha$,
the unnormalized confidence in $\mathcal{M}_\alpha$~is given by the ``Akaike
weight'' $\exp(-{\rm AIC}_\alpha/2)$. Informally, in a three-way comparison,
the relative probability that $\mathcal{M}_\alpha$~is statistically preferred is
\begin{equation}
\label{eq:lastAIC}
{\cal L}(\mathcal{M}_\alpha)=
\frac{\exp(-{\rm AIC}_\alpha/2)}
{\exp(-{\rm AIC}_1/2)+\exp(-{\rm AIC}_2/2)+\exp(-{\rm AIC}_3/2)}\;.
\end{equation}

\begin{figure}
\includegraphics[angle=0,scale=1.0]{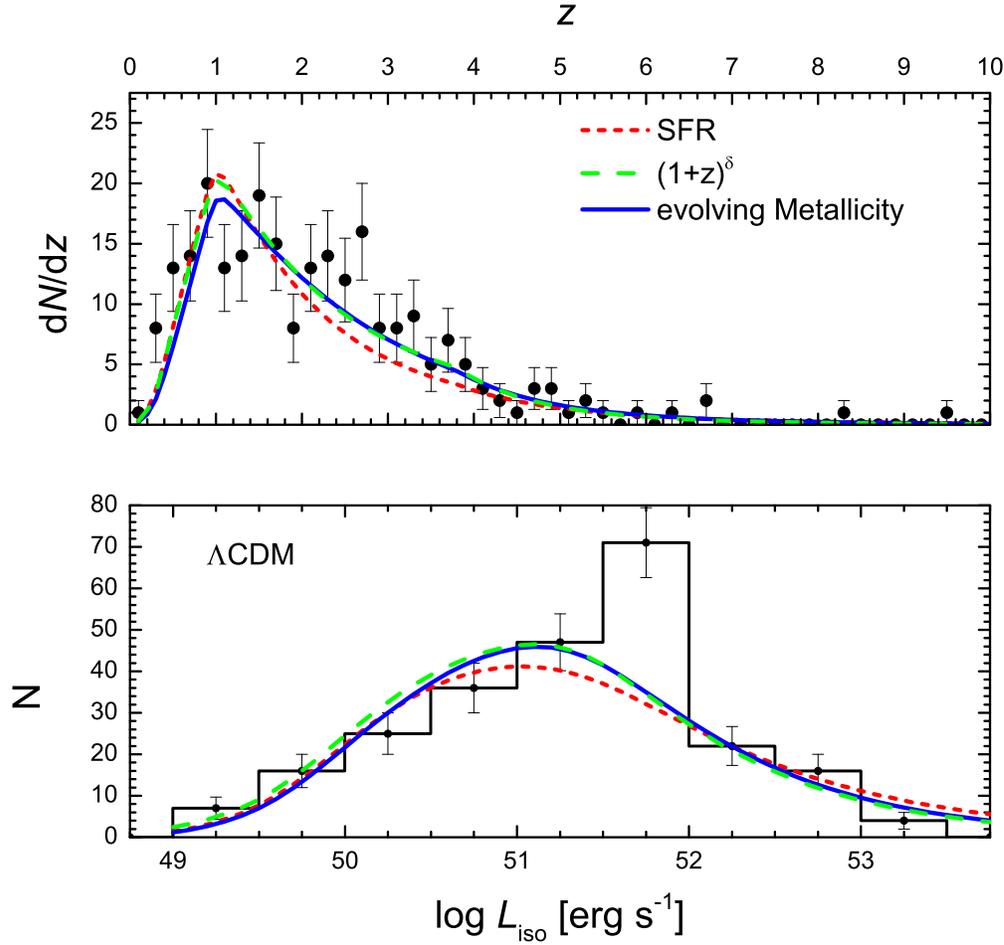}
\caption{Distributions in $z$ and $L$ of 244 \emph{Swift} GRBs with
firm redshift measurements in the $\Lambda$CDM cosmology (the solid points and steps,
with the number of GRBs in each bin indicated by a dark point with Poisson error bars).
The dotted lines (red) show the expected distribution for the case of no evolution.
The results of density and metallicity evolution models are shown with green dashed
lines and blue solid lines, respectively.}\label{f4}
\end{figure}

\begin{figure}
\includegraphics[angle=0,scale=1.0]{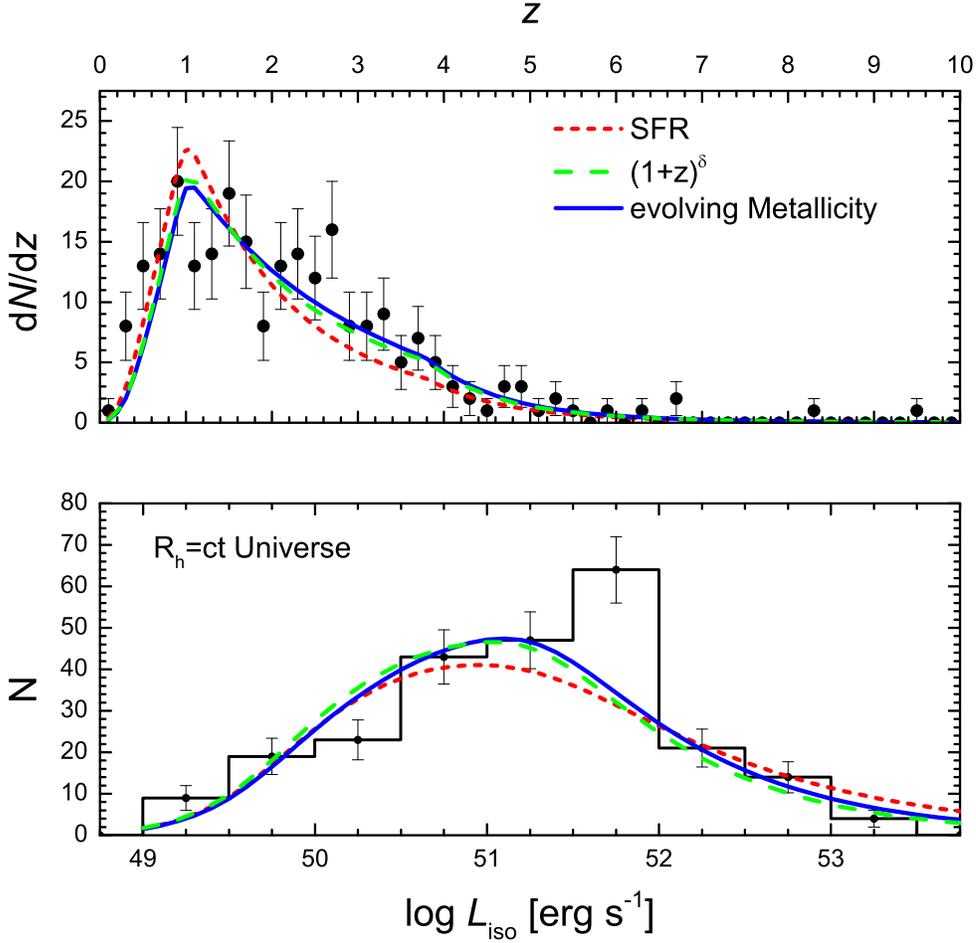}
\caption{Same as Figure~4, but for the $R_{\rm h}=ct$ Universe.}\label{f5}
\end{figure}

\subsubsection{No-Evolution Model}
This model is for the GRB rate that purely follows the SFR.
Figure~4 shows the $z$ and $L$ distributions of 244 \emph{Swift}
GRBS in the $\Lambda$CDM cosmology. If the function $f(z)$ is constant
(dotted red line), the expectation from the SFR alone (i.e., the non-evolution
case) does not provide a good representation of the observed $z$ and $L$
distributions of our sample. In particular, the rate of GRBs at high-\emph{z}
is under-predicted and the fit of the $L$ distribution is not as good as those of
the density evolution model or metallicity threshold model, more fully described
below. This is confirmed by a more detailed statistical analysis. Indeed, on the
basis of the AIC model selection criterion, we can discard this model as having
a likelihood of only $\sim0.2\;\%$ of being correct compared to
the other two $\Lambda$CDM models.

Figure~5 shows the redshift and luminosity distributions of 244 \emph{Swift}
GRBS in the $R_{\rm h}=ct$ Universe. The results of our fitting from the SFR alone
(i.e., with a constant $f[z]$) are indicated with dotted red lines, which again are
incompatible with the observations. On the basis of the AIC model selection
criterion, we can discard the no-evolution model as having a likelihood of only
$\sim0.1\;\%$ of being correct compared to the other two $R_{\rm h}=ct$ models.

\subsubsection{Density Evolution Model}
This model assumes that the GRB rate follows the SFR in conjunction with
an additional evolution characterized by $(1+z)^{\delta}$. In $\Lambda$CDM,
we find that $\delta=0.80$ reproduces the observed $z$ and $L$ distributions
(green dashed lines in Figure~4) quite well. In this model, the slope of the
high-\emph{z} SFR is characterized by an index $\alpha=-3.06_{-2.01}^{+2.01}$.
The range of high-\emph{z} SFH's with $\alpha\in(-5.07,-1.05)$ is marked with
a shaded band in figure~1, in comparison with the available data. It is
interesting to note that Wang (2013) derived a similar slope ($\alpha=-3.0$)
for the high-\emph{z} SFR. Wu et al. (2012) showed that the GRB formation rate
in $\Lambda$CDM decreases with a power index of $\sim-3.8$ for $z \ga 4$,
in good agreement with the SFR we derive here at the $1\sigma$ confidence
level. Using the AIC model selection criterion, we find that among the $\Lambda$CDM
models, this one is statistically preferred with a relative probability $\sim 57.3\;\%$.

In the $R_{\rm h}=ct$ Universe, we simultaneously fit the observed $z$ and $L$ distributions of
\emph{Swift} GRBs using $\dot{\rho}_{\rm GRB}(z)=k_{\rm GRB}R_{\rm SFR}(z)(1+z)^{\delta}$,
together with the piecewise smooth $R_{\rm SFR}(z)$ concatenated from equations~(\ref{SFRtwo})
and (\ref{HSFR}); the best fit corresponds to $\delta=1.03$. We find that a high-\emph{z} SFR
with slope $\alpha=-4.47_{-2.34}^{+2.30}$ is required to reproduce both the observed
$z$ and $L$ distributions (green dashed lines in figure~5). Again using the AIC
model selection criterion, we find that this model is somewhat disfavored statistically
compared to the other two $R_{\rm h}=ct$ models, with a relative probability
of $\sim 10.4\;\%$.

\subsubsection{Metallicity Evolution Model}
This model assumes that the GRB rate is proportional to the star formation history
with an additional evolution in cosmic metallicity (i.e., $f[z] \propto \Theta[\epsilon,z]$).
For $\Lambda$CDM, we find that a high-$z$ SFR with index $\alpha=-2.41_{-2.09}^{+1.87}$
and a metallicity evolution parameter $\epsilon=0.52$ fits the data best
(blue solid lines in Figure~4). The $\chi_{\rm dof}^{2}$ for this fit is
$56.0/42=1.33$. In general, fitting the observations with this model produces
better consistency than the non-evolution model. According to the AIC, the
metallicity evolution model in $\Lambda$CDM is slightly disfavored compared
to the more general density evolution model, but the differences are statistically insignificant
($\sim 42.5\;\%$ for the former versus $\sim 57.3\;\%$ for the latter).
We conclude that in the context of $\Lambda$CDM, the required density
evolution may be due to an evolving metallicity.

In the context of the $R_{\rm h}=ct$ Universe, the best fit is produced
with a high-$z$ SFR with index $\alpha=-3.60_{-2.45}^{+2.45}$ and a
metallicity evolution with $\epsilon=0.44$. The $\chi_{\rm dof}^{2}$
for this fit is $54.3/42=1.29$. This model is represented by the
blue solid lines in figure~5. The AIC shows that the likelihood of
this model being correct is $\sim 89.5\;\%$ compared to the other
two $R_{\rm h}=ct$ models examined above. Unlike the situation
with $\Lambda$CDM, here there is a clear indication that
abundance evolution is required to account for the SFR/GRB data.

\section{Discussion and Conclusions}

We have used the cumulative redshift distribution of the latest sample
of \emph{Swift} GRBs above a fixed luminosity limit, together with the
star formation history over the interval $z\in(0,4)$, to compare the
predictions of $\Lambda$CDM and the $R{\rm h}=ct$ Universe.
With $\Lambda$CDM as the background spacetime, earlier work had
already demonstrated that in this cosmology the SFR underproduces
the GRB rate density at high redshifts. It has been suggested that
this effect can be understood if a modest evolution, parameterized
as $f(z)=(1+z)^{0.80}$, is included; we have confirmed in both
$\Lambda$CDM and $R_{\rm h}=ct$ that this factor may be readily
explained as an evolution in metallicity. However, we have also
found that a comparison with the observational data shows that a
relatively high metallicity cut ($Z=0.52Z_{\odot}$ in $\Lambda$CDM
and $Z=0.44Z_{\odot}$ in $R_{\rm h}=ct$) is required, in contrast to
previous work that suggested LGRBs occur preferentially
in low metallicity environments, i.e., $Z\sim0.1-0.3Z_{\odot}$.

For both cosmologies, we have shown that if these results are correct,
then by assuming that such trends continue beyond $z\simeq4$, the
adoption of a simple power-law approximation for the high-\emph{z}
($\ga 3.8$) SFR , i.e., $R_{\rm SF}\propto[(1+z)/4.8]^{\alpha}$,
we may also constrain the slope $\alpha$ using the GRB data. We
have found for $\Lambda$CDM that the SFR at $z \ga 3.8$ shows a
decay with slope $\alpha=-2.41_{-2.09}^{+1.87}$. And using a
simple relationship between the GRB rate density and the SFR,
including an evolution in metallicity, we have demonstrated that
the $z$ and $L$ distributions of 244 \emph{Swift} GRBs can be well
fitted by our updated SFH, using a threshold in the metallicity for
GRB production.

The best fit for the redshift distribution of the \emph{Swift} GRBs in the
$R_{\rm h}=ct$ Universe requires a slightly different rate than that in
$\Lambda$CDM, though still with an additional evolutionary effect,
which could be a high metallicity cut of $Z=0.44Z_{\odot}$. Assuming
that the GRB rate is related to the SFR with this evolving metallicity, we have
found that in the $R_{\rm h}=ct$ Universe the slope of the high-\emph{z} SFR
would be $\alpha=-3.60_{-2.45}^{+2.45}$.

The principal goal of this work has been to directly compare the predictions
of $\Lambda$CDM and $R_{\rm h}=ct$ and their ability to account for the
GRB/SFR observations. Aside from the issue of whether or not the GRB-redshift
distribution is consistent with the SFR in either model, we have also examined
which of these two cosmologies fits the data better, and is therefore
statistically preferred by the Aikake Information Criterion in a one-on-one
comparison.

To keep the complexity of this problem manageable, we have chosen to find
the best fits to the data by optimizing four free parameters ($\alpha$, $k_{\rm GRB}$,
$L_{\star}$ and $a_{L}$), though the models themselves were held fixed by the
concordance values of $H_0$, $\Omega_m$, $\Omega_\Lambda$ and the
dark-energy equation-of-state in the case of $\Lambda$CDM, and the same
value of $H_0$ for $R_{\rm h}=ct$. The two models produce very similar
profiles in the distance-redshift relationship (Melia 2012a; Wei et al. 2013),
so it is not very surprising to see that both can account quite well for the
observed SFR-GRB rate correlation.

However, the AIC does not favor these models equally. From Table 2,
we find that a direct comparison between the best $\Lambda$CDM fit (entry 2
in this table) and the best $R_{\rm h}=ct$ fit (entry 6) favors the latter
with a relative probability $\sim63.4\;\%$ versus $\sim 36.6\;\%$ for the
standard model. If we further assume that the required evolutionary
effect is indeed due to changes in metallicity, so that we now compare
entries 3 and 6 in Table 2, then the AIC favors $R_{\rm h}=ct$ with
a relative probability $\sim 70.0\;\%$ versus $\sim 30.0\;\%$ for
$\Lambda$CDM. However, if the required evolutinary effect is
simply due to density and not changes in metallicity (entries 2 and
5), the AIC favors $\Lambda$CDM with a relative probability of
$\sim 83.2\;\%$ versus $\sim 16.8\;\%$ for $R_{\rm h}=ct$.

The statistical significance of these likelihoods has been
investigated theoretically, e.g., by Yanagihara \& Ohmoto
(2005). Its variability has also been studied empirically; for example, by
repeatedly comparing $\Lambda$CDM to other cosmological models on the basis
of data sets generated by a bootstrap method (Tan \& Biswas 2012). It is
known that the AIC is increasingly accurate when the number of data points
is large, but it is felt that in all cases, the magnitude of the difference
$\Delta=\allowbreak {\rm AIC}_2 -\nobreak {\rm AIC}_1$ should provide a
numerical assessment of the evidence that model~1 is to be preferred over
model~2. A~rule of thumb that has been used in the literature is that if
$\Delta\la3$, it is mildly strong; and if $\Delta\ga5$, it is quite strong.

Therefore, our conclusion from the comparative study we have reported here is
that---based on the currently available GRB/SFR observations---the
$R_{\rm h}=ct$ Universe is mildly favored over $\Lambda$CDM in a one-on-one
comparison if the required evolution is due to changes in metallicity (for
which $\Delta\approx 1.7$). However, $\Lambda$CDM is mildly favored over
$R_{\rm h}=ct$ (with $\Delta\approx 3.2$) if instead the evolution
is with density.

The prevailing view at the moment seems to be that changes in
metallicity are responsible for the required evolution so, in this
context, the GRB/SFR data tend to be more consistent with the
predictions of $R_{\rm h}=ct$ than those of the concordance model.
Note that the likelihood estimates we have made here were based on
the use of priors for $\Lambda$CDM. Were we to optimize $H_0$
along with the other four parameters (for both models), and
$\Omega_m$, $\Omega_\Lambda$ and the dark-energy equation of state
for $\Lambda$CDM, we could certainly lower their $\chi^2$ for the best
fits, but the AIC strongly penalizes models with many free parameters.
The $\chi^2$ values listed for $\Lambda$CDM in Table 2 would need to
decrease by at least 6 in order to compensate for the increase due to
the factor $2k$ in the expression ${\rm AIC}=\chi^2+2k$. This seems
unlikely since the fits using the concordance model are already
rather good.

Refinements in future measurements of the GRB rate and SFR may show that
the currently believed explanation for their differences (i.e., an evolution
in metallicity) is incorrect. In that case, a reassessment of these
comparisons may produce different results. As of now, however, it appears
that the SFR underproduces the observed GRB rate unless some additional
evolution were present to broaden their disparity with increasing redshift.
We have found that such an evolution is consistent with a relatively high
metallicity cutoff for the LGRBs.

\section*{Acknowledgments}
We thank X. H. Cui, X. Kang, E. W. Liang, and F. Y. Wang for helpful discussions.
This work is partially supported by the National Basic Research Program (``973" Program) of China
(Grants 2014CB845800 and 2013CB834900), the National Natural Science Foundation of China
(grants Nos. 10921063, 11273063, 11322328, and 11373068), the One-Hundred-Talents Program
and the Youth Innovation Promotion Association of the Chinese Academy of Sciences, and
the Natural Science Foundation of Jiangsu Province. F.M. is grateful to Amherst College for its support
through a John Woodruff Simpson Lectureship, and to Purple Mountain Observatory
in Nanjing, China, for its hospitality while this work was being carried out.
This work was partially supported by grant 2012T1J0011 from The Chinese Academy of
Sciences Visiting Professorships for Senior International Scientists, and grant
GDJ20120491013 from the Chinese State Administration of Foreign Experts Affairs.
We also thank the anonymous referee for providing many comments and suggestions
that have led to a significant improvement in the presentation of the material
in this paper.

\end{document}